\RequirePackage{fix-cm}

\documentclass[12pt, Numbered]{elsarticle}

\usepackage[utf8]{inputenc}
\usepackage[T1]{fontenc}
\usepackage{lmodern}
\usepackage{geometry}

\usepackage{latexsym}
\usepackage{graphicx}
\usepackage{dialogue}
\usepackage{tikz}
\usetikzlibrary{decorations.pathreplacing}
\usepackage{arydshln}
\usepackage{amssymb}
\usepackage{amsfonts}
\usepackage{amsmath}
\usepackage{dsfont}
\usepackage{pifont}

\usetikzlibrary{arrows,decorations,shapes,automata,positioning,decorations.pathmorphing}
\tikzset{
  every picture/.style = {
    thick,
    >=stealth',
    node distance = 1.5em and 3em,
  }
  ,
  cross line/.style = {
    preaction = {
      draw=white,
      -,
      line width=4pt
    }
  }
  ,
  state/.style = {
    rectangle,
    rounded corners = 5pt,
    font = \footnotesize,
    draw,
    minimum width = 1em,
    minimum height = 1em
  }
  , 
  label-state/.style = {
    sloped,
    font = \scriptsize,
    label distance = -2pt
  }
  , 
  label-edge/.style = {
    font = \scriptsize,    
    label distance = -2pt
  }
}
\usepackage{caption}
\usepackage{float}
\usepackage[shortlabels]{enumitem}
\usepackage{xspace}
\usepackage{rotating}
\usepackage{tabularx}
\usepackage{longtable,tabu}
\usepackage{xcolor,colortbl}
\definecolor{kugray5}{RGB}{224,224,224}
\usepackage{multirow}
\usepackage{multicol}
\usepackage{hhline}
\usepackage{xcolor}
\usepackage{tabularray}
\usepackage{float}

\NewTableCommand{\mcc}[1][1]{\multicolumn{#1}{c}}
\usepackage{textcomp}
\usepackage{amsmath}
\usepackage{array, multirow, bigdelim, makecell, booktabs} 

\usepackage{subfig}

\newcommand{\Logic}{$\mathcal{L}_{(\mathsf{intel})}$\xspace}
\newcommand{\Form}{$\mathcal{F}_{(\mathsf{intel})}$\xspace}

\newcommand{\C}{\texttt{\textit{\textbf{C}}}}
\newcommand{\R}{\texttt{\textit{\textbf{R}}}}

\usepackage{booktabs}
\usepackage{tikz}
\usetikzlibrary{decorations.pathreplacing}

\usepackage{hyperref}

\begin{document}
\begin{frontmatter}



\title{A Dynamic Logic for Information Evaluation in Intelligence}


\author{Benjamin Icard}

\address{LIP6, Sorbonne Université, Institut Jean-Nicod (CNRS, ENS-PSL, EHESS), Paris, France}

\begin{abstract}

In the field of human intelligence, officers use an alphanumeric scale, known as the Admiralty System, to rate the credibility of messages and the reliability of their sources \cite{NATOAJP-2.1}. During this evaluation, they are expected to estimate the credibility and reliability dimensions independently of each other \cite{STANAG2003}. However, empirical results show that officers perceive these dimensions as strongly correlated \cite{Baker&al1968}. More precisely, they consider credibility as playing the leading role over reliability, the importance of which is only secondary \citep{Samet1975}. In this paper, we present a formal evaluative procedure, called \Logic, in line with these findings. We adapt dynamic belief revision to make credibility the main dimension of evaluation and introduce dynamic operators to update credibility ratings with the source's reliability. In addition to being empirically sound, we show that \Logic provides an effective procedure to classify intelligence messages along the descriptive taxonomy presented in \cite{Icard2023a}.

\end{abstract}

\begin{highlights}
\item Contrary to some crucial assumption of the intelligence doctrine (e.g. NATO STA\-NAG-2511, 2003), results from experiments have shown that intelligence officers perceive credibility as a more important dimension then reliability during information evaluation. Results also show that officers fail to distinguish facts from interpretations with the extant procedure, contrary to another central assumption.

\item We define a dynamic belief revision logic \Logic for intelligence scoring in line with those findings. Our framework relies on doxastic plaubility models to make credibility play the leading role during information evaluation, and introduces dynamic operators to make reliability an adjustement of credibility.

\item While keeping assumptions at a minimum, \Logic remains valid when we consider new empirical observations. First designed to address issues on credibility versus reliability, \Logic also fits with findings on the facts versus interpretations issues, and is consistent with solutions proposed in that respect.

\item With such an integration, the scope of \Logic extends to other intelligence tasks. The setting is evaluative by design but can also serve more descriptive purposes. In line with the need to separate facts from interpretations, \Logic proves to be an effective procedure to classify intelligence messages into descriptive categories. 

\end{highlights}

\begin{keyword}
Human Intelligence \sep Information Evaluation \sep Admiralty Scale \sep Dynamic Belief Revision Theory \sep Scoring \sep Informational Taxonomy \sep Intelligence Processing

\end{keyword}

\end{frontmatter}

\section{Introduction}
\label{sec:intro}


In the field of human intelligence (HUMINT), evaluating information is a crucial step of information processing. According to doctrinal texts \citep[e.g.][]{STANAG2003,NATOAJP-2.1}, intelligence officers use a 6$\times$6 Alphanumeric scale called the Admiralty System, or NATO System, to assess the credibility of message contents and the reliability of their sources. During this procedure, officers are required to respect two methodological recommendations in particular \citep[see][p. 2]{STANAG2003}. One of those recommendations, hereafter called the ``\textit{facts versus interpretations recommendation}'', is to always distinguish objective facts from their subjective interpretations. Another recommendation, hereafter called the ``\textit{credibility versus reliability recommendation}'', asks officers to evaluate the credibility and reliability dimensions on an independent non-overlapping basis.

\medskip
This evaluative procedure has been challenged by experimental results. The scale suffers from various limitations which make officers unable to respect either recommendation. On the one hand, officers fail to respect the \textit{facts versus interpretations recommendation} since the extant 6$\times$6 scale, which is evaluative by nature, cannot be used to identify more objective, i.e. descriptive, facts hidden behind their subjective interpretations of intelligence. On the other hand, officers are unable to respect the \textit{credibility versus reliability recommendation} either. In fact, officers perceive dimensions of credibility and reliability as dependent and strongly correlated. More precisely, they perceive credibility as the prevalent evaluative dimension while reliability only plays a secondary role, to adjust assessments of credibility. 
\medskip

In this paper, we propose a logical framework to overcome issues regarding \textit{credibility versus reliability}. We define a dynamic logic for belief revision, called \Logic, in which credibility is the main dimension of the setting. Plausibility models help express prior credibility distributions on messages through the definition of various degrees of credibility strength. The source's reliability comes to update those prior distributions through the definition of varous dynamic operators. To keep theoretical assumptions at a minimum, we do not address issues regarding \textit{facts versus interpretations} directly. Extant proposals to enforce this latter recommendation, in particular the one by \cite{Icard2023a}, rely on the assumption that field officers only use a 3$\times$3 subpart of the 6$\times$6 scale by grouping the credibility and reliability dimensions into three categories instead of six \citep[e.g.][]{Teigen&Brun1995,mandel2022meta}. Nonetheless, we argue that our logical setting remains consistent with this 3$\times$3 subscale. We show that \Logic leads to retrieve the 3$\times$3 descriptive taxonomy of intelligence messages proposed in \cite{Icard2023a}.

\medskip
 In section \ref{sec:infoeval}, we present the procedure traditionnally used to evaluate information in human intelligence (HUMINT). We first describe the $6\times6$ Admiralty System using credibility and reliability dimensions for that purpose, and then the two guidelines provided to help officers manipulating the scale correctly (subsection \ref{ssec:alpha1}). 
 Experiments show that the scale dimensions are misunderstood by officers and that none of two recommendations are respected by them \citep[e.g.][]{Baker&al1968,Samet1975,mandel2022meta} (subsection \ref{ssec:strength/short}). 

\medskip

Section \ref{sec:issues} reviews extant solutions to overcome the procedure limitations. We start by presenting informal approaches intended to clarify the scale dimensions and to help officers comply with the \textit{facts versus interpretations recommendation} (\ref{ssec:informalicard}). In this regard, we put particular emphasis on the descriptive taxonomy of intelligence messages introduced by \cite{Icard2023a}, and show how the categories, or messsage types, of this taxonomy can be ordered qualitatively from best to worst. Then, we consider formal proposals to specify the scale dimensions and to address issues regarding the \textit{credibility versus reliability recommendation}, based on quantitative approaches (probabilistic reasoning), qualitative methods (non-classical logics), or mixed approaches (subjective logic) (subsection \ref{ssec:extantformal}). Finally, we briefly present our new formal logic \Logic for information evaluation and the constraints it should respect to comply with past empirical observations (subsection \ref{ssec:proposal}).

\medskip
In section \ref{sec:new}, we describe our evaluative setting \Logic in detail. Adaptating previous works by \citep{Aucher2004,Aucher2008,vanDitmarsch2005,VanDitmarsch&Labuschagne2007}, the logic relies on numerical belief revision to make credibility the main dimension of evaluation and reliability play an adjustement role (subsection \ref{ssec:formal}). Credibility ratings are captured through a set of \textit{conditional credibility operators} indexed from 1 to 6 (subsection \ref{ssec:cred}). Reliability ratings are captured through \textit{updates of credibility degrees} labeled from A to F (subsection \ref{ssec:rel}). Once applied, these updates provide \textit{posterior} credibility scores to messages based on the reliability of their source.     

\medskip
Section \ref{sec:classifying} shows the extension of \Logic to the categorization of intelligence messages. In this section, we make extensive use of the grouping hypothesis according to which officers group the credibility and reliability dimensions into 3$\times$3 instead of 6$\times$6 dimensions \cite{mandel2022meta}. We first compute the landscape of posterior scores \textit{without the hypothesis}, that is to say: by applying \Logic directly to the 6$\times$6 alphanumeric scale (subsection \ref{ssec:scoring}). We then \textit{endorse the grouping hypothesis} to see how this landscape of scores evolves in the 3$\times$3 case (subsection \ref{ssec:meanscores}). Defining rules for \textit{grouped} reliability updates, we compute 9 posterior credibility scores for messages and rank them quantitatively from highest to lowest.\footnote{In the same way message types are ranked qualitatively from best to worst in subsection \ref{ssec:informalicard}.} The quantitative ordering we obtain is congruent with the qualitative ordering obtained before, so that, in the 3$\times$3 configuration, the evaluative approach of \Logic matches with the descriptive approach of \cite{Icard2023a}. In other words, \Logic helps retrieve the descriptive categories of this approach, and gives a procedure to categorize intelligence messages more objectively (subsection \ref{ssec:match}). 

 \medskip
In section \ref{sec:conclusion}, we conclude by summarizing our proposal for information evaluation in intelligence. We briefly highlight the main strengths and inner specificities of our formal setting. Finally, we emphasize the various aspects on which our approach could be extended in the future, in particular to other intelligence tasks.

\clearpage
\section{Information Evaluation in Intelligence}
\label{sec:infoeval}














\subsection{The Evaluative Procedure According to the Intelligence Doctrine}
\label{ssec:alpha1}


To perform the task of information evaluation, officers rely a 6$\times$6 scale provided with six levels for evaluating the credibility of message contents as well as six levels for evaluating the reliability of their sources \citep[see][]{STANAG2003,NATOAJP-2.1}. Both credibility and reliability ratings are then crossed to provide intelligence messages with global evaluations.

\begin{table}[H]

\scriptsize
\centering

\renewcommand{\arraystretch}{1.5}
\begin{tabular} {|c|c|c|c|c|c|c|c|} 

 \hline
 
\multicolumn{2}{|c|}{\cellcolor{kugray5}\textbf{Message}} & \multicolumn{6}{c|}{\textit{\textbf{Content Credibility}}}\\

    \cline{3-8}
    \multicolumn{2}{|c|}{\cellcolor{kugray5}\textbf{Scores}} &\textbf{1}&\textbf{2}&\textbf{3}&\textbf{4}&\textbf{5}&\textbf{6}\\
    \hline
    \multirow{6}{*}{\begin{turn}{90}\textit{\textbf{Source Reliability}}\ \ \end{turn}}& \textbf{A}&\textbf{A1}&\textbf{A2}&\textbf{A3}&\textbf{A4}&\textbf{A5}&\textbf{A6}\\
    \cline{2-8}
                  & \textbf{B}&\textbf{B1}&\textbf{B2}&\textbf{B3}&\textbf{B4}&\textbf{B5}&\textbf{B6}\\
    
       \cline{2-8}
                       & \textbf{C}&\textbf{C1}&\textbf{C2}&\textbf{C3}&\textbf{C4}&\textbf{C5}&\textbf{C6}\\
    
       \cline{2-8}
                      & \textbf{D}&\textbf{D1}&\textbf{D2}&\textbf{D3}&\textbf{D4}&\textbf{D5}&\textbf{D6}\\
    
       \cline{2-8}
                     & \textbf{E}&\textbf{E1}&\textbf{E2}&\textbf{E3}&\textbf{E4}&\textbf{E5}&\textbf{E6}\\
    
       \cline{2-8}
                      & \textbf{F}&\textbf{F1}&\textbf{F2}&\textbf{F3}&\textbf{F4}&\textbf{F5}&\textbf{F6}\\
      
    \hline

\end{tabular}
\quad
\renewcommand{\arraystretch}{1.4}
\begin{tabular}{|l|l|}
    
    \hline
   
    \rowcolor{kugray5}\textbf{\textit{Content Credibility}}  & \textbf{\textit{Source Reliability}} \\
   
    \hline
    
      \textbf{1:} Confirmed & \textbf{A:} Completely Reliable \\\hline

    \textbf{2:} Probably True & \textbf{B:} Usually Reliable \\\hline

    \textbf{3:} Possibly True & \textbf{C:} Fairly Reliable \\\hline

    \textbf{4:} Doubtfully True & \textbf{D:} Not Usually Reliable \\\hline

    \textbf{5:} Improbable & \textbf{E:} Unreliable \\\hline

    \textbf{6:} Cannot Be Judged & \textbf{F:} Cannot Be Judged 
      \\ \hline

    \end{tabular}

\caption{The Traditional Admiralty System for Information Evaluation: the 6$\times$6 Alphanumeric Scale (left), and the labels corresponding to the various levels of credibility and reliability (right).\label{tab:scale}}

\end{table}

\begin{table}[h]
  \centering

\small
\begin{tabular}{|c|c|p {8,5cm}|}
    \hline
     \rowcolor{kugray5}\textbf{Ratings} & \textbf{Linguistic Labels} & \multicolumn{1}{|c|}{\textbf{Descriptions}} \\
    \hline
    
     \textbf{1}  & \small{Confirmed} & \small{Confirmed by other independent sources; consistent with other information on the subject.} 
      \\\hline
    
     \textbf{2} & \small{Probably True} &  \small{Not confirmed; consistent with other information on the subject.} 
     \\\hline
    
      \textbf{3}  & \small{Possibly True} &  \small{Not confirmed; agrees with some other information on the subject.} 
     \\\hline

      \textbf{4} & \small{Doubtfully True} &  \small{Not confirmed; possible; no other information on the subject.} 
     \\\hline

      \textbf{5} & \small{Improbable} &  \small{Not confirmed; contradicted by other information on the subject.}  
     \\\hline
 
      \textbf{6} & \small{Cannot Be Judged} & \small{No basis exists for evaluating the validity of the information.}
    \\ \hline
    \end{tabular}
\caption{The Credibility of the Message Content.\label{tab:cred}}
\end{table}%

Tables \ref{tab:cred} and \ref{tab:rel} provide the descriptions of the credibility and reliability dimensions respectively. According to the phrasings of those descriptions, credibility and reliability are intended to help officers  give subjective estimations of the more descriptive dimensions of truth (for message contents) and of honesty (for message sources). 

\medskip
Basically, \textit{credibility ratings} are captured through 6 degrees ranging from \textbf{1} to \textbf{6} (see Table \ref{tab:cred}). Those degrees express \textit{decreasing levels of confirmation} given some \textit{contextual evidence} the officer has \textit{for}, or \textit{against}, the message being true. Evidence is \textit{consistent} with the message in case some pieces of evidence are true when the message is true. Evidence is \textit{inconsistent} with the message when pieces of evidence are true but the message is false. Finally, precondition for evaluating the message fails when evidence is missing for assessing the truth or falsity of the message. Accordingly, degree \textbf{1} corresponds to cross-checked certainty: all the evidence the officer has is consistent with the message being true (``Confirmed''). Degrees \textbf{2} to \textbf{5} correspond to adverbial modulations that express weaker states of consistency: from \textit{high consistency} (``Probably True'') to \textit{moderate consistency} (``Possibly True''), \textit{weak consistency} (``Doubtfully True'') and \textit{blatant inconsistency} (``Improbable''). Degree \textbf{6} is ascribed when precondition fails for making an evaluation: \textit{no evidence} exists for assessing the credibility of the message (``Cannot Be Judged''). 

\medskip
\begin{table}[h]
\small
  \centering
\begin{tabular}{|c|c|p {8,5cm}|}
    \hline
     \rowcolor{kugray5}\textbf{Ratings} & \textbf{Linguistic Labels} & \multicolumn{1}{|c|}{\textbf{Descriptions}} \\
    \hline

     \textbf{A}  & \small{Completely Reliable} & \small{No doubt of authenticity, trustworthiness, or competency; has a history of complete reliability.} 
\\\hline

     \textbf{B} & \small{Usually Reliable} & \small{Minor doubt about authenticity, trustworthiness, or competency; has a history of valid information most of the time.} 
     \\\hline
     
      \textbf{C}  & \small{Fairly Reliable} & \small{Doubt of authenticity, trustworthiness, or competency but has provided valid information in the past.} 
\\\hline

      \textbf{D} & \small{Not Usually Reliable} & \small{Significant doubt about authenticity, trustworthiness, or competency but has provided valid information in the past.} 
\\\hline

      \textbf{E} & \small{Unreliable} & \small{Lacking in authenticity, trustworthiness, and competency; history of invalid information.}
\\\hline

      \textbf{F} & \small{Cannot Be Judged} & \small{No basis exists for evaluating the reliability of the source.}\\ \hline

    \end{tabular}%
\caption{The Reliability of the Message Source.\label{tab:rel}}
\end{table}%

\medskip
Reliability ratings are captured through six levels ranging from \textbf{A} to \textbf{F} (see Table \ref{tab:rel}). These ratings express \textit{decreasing degrees of trustworthiness} reflecting the officer's trusting attitude regarding the source's honesty. Level \textbf{A} corresponds to absence of suspicion (``Completely Reliable''): \textit{no doubt of authenticity}. Levels from \textbf{B} to \textbf{E} correspond to increasing suspicion against the source's honesty: from \textit{minor doubt} (``Usually Reliable'') to \textit{doubt} (``Fairly Reliable''), \textit{significant doubt} (``Not Usually Reliable'') and \textit{lack of trustworthiness} (``Unreliable''). Here again, level \textbf{F} is ascribed  when precondition fails for making an evaluation: \textit{no evidence} exists for assessing the reliability of the source (``Cannot Be Judged''), — either because the source is new to the officer or because she has not delivered relevant information in the past.  
\medskip




\medskip
Moreover, the STANAG doctrine also provides officers with methodological recommendations for using the 6$\times$6 alphanumeric scale properly. For convienience, we choose to call them the \textit{facts versus interpretations recommendation} and the \textit{credibility versus reliability recommendation}, respectively \citep[see][A-2]{STANAG2003}. The \textit{facts versus interpretations recommendation} is expressed as follows in the doctrine:  

\begin{quote}
\textbf{Facts versus Interpretations.} ``\textit{Intelligence reports transmit facts and/or assessments. The distinction between fact and interpretation must always be clearly indicated.}''
\end{quote}

According to this guideline, officers should always clearly distinguish intelligence facts from the subjective interpretations they make of those facts when rating intelligence. From a more epistemological perspective, as it has been noticed in e.g. \cite{Icard2023a}, this distinction between facts and interpretations is analogous to the descriptive versus evaluative distinction with regard to information. When using the credibility and reliability dimensions to appreciate pieces of information, officers have an interpretative perspective that is of an \textit{evaluative} nature. On the side of facts, now, officers attempt to consider intelligence messages from an objective point of view, independently, or at least separately, from their subjective evaluation of them.      

\medskip
In this article, we pay more attention to the \textit{credibility versus reliability recommendation}, before returning to the \textit{facts versus interpretations recommendation} as a second step.  The \textit{credibility versus reliability recommendation} is expressed as follows in the doctrine \citep[see again][A-2]{STANAG2003}:     

\begin{quote}
\textbf{Credibility versus Reliability.} ``\textit{Reliability and credibility, the two aspects of evaluation, must be considered independently of each other}''. 
\end{quote}

This recommendation states that intelligence officers should perform their evaluations of the credibility and reliability dimensions on an independent basis. According to the Field Manual 30-5: ``\textit{Although both letters and numerals are used to indicate the evaluation of an item of information, they are independent of each other}'' \citep{FM30-5}. Evaluations of credibility should not interact with evaluations of reliability and, conversely, evaluations of reliability should not interplay with evaluations of credibility. But this recommandation presupposes that the credibility and reliability dimensions are perceived as independent and non-overlapping by officers. We will see that this is not the case. 
\medskip

Both the \textit{facts versus interpretations} and the \textit{credibility versus reliability recommendations} are in fact challenged by empirical observations. Results show that intrinsic limitations in the scale features lead to inconsistencies within and between raters that prevent them from respecting those two recommendations, only \textit{de facto} for the \textit{facts versus interpretations recommendation} but also \textit{de jure} for the \textit{credibility versus reliability recommendation}. 

\subsection{The Procedure's Shortcomings and Consequences for Officers}
\label{ssec:strength/short}



\subsubsection{Officers' Misunderstandings of the Evaluative Dimensions}
\label{ssec:lackof}

Let us start by pointing out some of the strengths of the Admiralty scale. First of all, credibility and reliability are relevant dimensions for information evaluation in general and in the intelligence domain, more specifically. The scale for information evaluation in intelligence is quite sensitive: officers are provided with five levels for evaluating both dimensions, as well as one extra-dimension when no evaluation is possible. Another positive aspect is that methodological guidelines (i.e. \textit{facts versus interpretation}, \textit{credibility versus reliability}) are provided to officers for them to use the alphanumeric scale properly. However, we will see now that the procedure suffers from various drawbacks that have been pointed out experimentally. Within and between raters inconsistencies show that officers do not have a clear understanding of the scale dimensions.\footnote{See \cite{icardthesis2019} for a more detailed review of those experimental results.}

\medskip

Concerning \textit{within-raters inconsistencies}, it turns out that officers do not draw a clear distinction between the credibility and reliability dimensions of the scale. When looking at 695 resultant scores provided by American army officers, \cite{Baker&al1968} observed that 87\% of the scores ($N$ = 608 out of 695) fell stricly along the diagonal of the scale, that is on the continuum \textbf{A1}-\textbf{B2}-\textbf{C3}-\textbf{D4}-\textbf{E5}-\textbf{F6}. This observation, also made by \cite{Samet1975}, reveals that officers, despite not considering credibility and reliability as strictly similar dimensions, perceive them as strongly correlated and tend to mix them up. Officers are confused about the inner meaning of the scale rating labels, their respective levels, and resultant scores. 
\medskip

Concerning \textit{between-raters inconsistencies}, \cite{Baker&al1968} showed that collected ratings differed from the instructor’s solution about 49\% of the time for credibility, and 15\% of the time for reliability. Similar high variability was observed regarding proposals asking for the substitution of absolute probability degrees to extant avderbial labels \citep[see e.g.][]{Kent1964}. For instance, \cite{Wark1964} observed only 53\% of inter-agreement that the modal adverb \textit{“possibly”} used in credibility rating \textbf{3} (``Cannot Be Judged'') correspond to a probability degree of 0.50 while \cite{Johnson1973} observed that the root adjective \textit{“possible”} was associated to a mean probability of 0.62 but with results varying from 0.04 to 0.80 across officers. Wark observed strong consensus that the adverb \textit{“probably”} was associated to a probability degree of 0.75 (90\% of inter-agreement) but this result was not replicated by Johnson for the root adjective “probable” which received a probability mean of 0.51 but with results varying from 0.10 to 0.99 across officers. Johnson also tested the adjective \textit{“improbable”} used in credibility rating \textbf{5}: it was assigned an absolute degree of 0.17 but with results varying from 0 to 0.70.\footnote{This lack of consistency in the interpretation of adverbial quantifiers has been observed in many other fields, such as linguistics (see \citealp{Lichtenstein&Newman1967} and \citealp{Budescu&Wallsten1985} on high variability for \textit{“possible”}) or medicine (compare the variability observed for \textit{“possible”} between \citealp{Obrien1989} and \citealp{Bryant&Norman1980}).} 

\medskip
This lack of consistent understanding results in the following issue: officers do not exploit the full possibilities of the 6$\times$6 matrix but only a subpart of it. In addition to that, they do not respect the two recommendations provided with the 6$\times$6 alphanumeric scale. On the one hand, officers cannot follow the \textit{facts versus interpretations} recommendation \textit{de facto}, based on the strict evaluative nature of the scale. On the other hand, they are unable to respect the \textit{credibility versus reliability} recommendation due to the overlapping aspect of those two dimensions in practice. Let us now review experimental results highlighting those failures and present extant proposals to overcome them.   










\subsubsection{The 6$\times$6 Scale is Not Used at Full Range by Officers}  

\label{sssec:notused}


The first consequence of the inconsistencies we have reviewed is that scale is not used at full range by officers. This can be observed at the level of resultant scores but also at the level of the credibility and reliability dimensions themselves.

\medskip
Concerning resultant scores, it turns out that officers only use a restricted subset of the set of 36 available ratings. In 1968, Baker, McKendry \& Mace analyzed 695 joint ratings obtained from two US intelligence corps during field exercises (Raw Number $N$ = 716) \citep[see][]{Baker&al1968}. When looking at the distribution of the ratings, Baker \& al. observed that 87\% of the scores \textit{fell stricly along the diagonal} of the scale, that is on the continuum \textbf{A1}-\textbf{B2}-\textbf{C3}-\textbf{D4}-\textbf{E5}-\textbf{F6} ($N$ = 608 out of 695). Moreover, score \textbf{B2} alone  comprised 75\% of all the ratings ($N$ = 518). In addition to that, 11\% of the scores fell (stricly) above the diagonal of the scale ($N$ = 75) whereas only 2\% of the scores fell (strictly) below the diagonal ($N$ = 12). 58\% of total possible scores were in fact never used by officers in Baker \& al' experiments ($N$ = 21 out of 36).\footnote{The scores that were never elicited by officers during those experiments are the following ones: from \textbf{A4} to \textbf{A6}, \textbf{B5} and \textbf{B6}, \textbf{C6}, \textbf{D5} and \textbf{D6}, \textbf{C1} and \textbf{C2}, \textbf{D1} to \textbf{D3}, \textbf{E1} to \textbf{E4}, \textbf{F1} to \textbf{F5}.}

\medskip
Concerning credibility and reliability, 
intelligence raters tend to group dimensions into three levels instead of six, and thus to use a 3$\times$3 matrix instead of a 6$\times$6 matrix. As observed in different experimental studies \citep[e.g][]{Teigen&Brun1995,teigen&brun1999directionality,budescu2003predicting,irwin2019improving}, adverbial quantifiers tend to be perceived as having either a positive or a negative directionality, and to be grouped together when they have the same directionality. In particular, Teigen \& Brun ran a study with a total of 62 students (32 American, 30 Norwegian) to determine whether 24 verbal probability expressions directed the students’ foci of attention toward a positive or a negative evaluation. Amongst those expressions, some were identical, or very close to, the verbal expressions used in the intelligence ratings: e.g. \textit{“absolutely certain”}, \textit{“probable”}, \textit{“possible”}, \textit{“doubtful”}, \textit{“improbable”}. In line with intuitions, it turned out that 100\% of the subjects that were tested for expressions \textit{“absolutely certain”}, \textit{“probable”} and \textit{“possible”} rated those expressions as positively directional, while 100\% of the subjects that were tested for \textit{“doubtful”} and \textit{“improbable”} rated them as negatively directional \citep[see][study 1 for details]{Teigen&Brun1995}.
\medskip

Recently, \cite{mandel2022meta} devoted two experiments to the phenomenon of directionality in the context of information evaluation. In line with Teigen and Brun’s preliminary results \citep{Teigen&Brun1995}, they observed that credibility ratings from \textbf{1} to \textbf{5} and reliability ratings from \textbf{A} to \textbf{E} can be separated into positive versus negative groupings. Graphically, indeed, a clear inflection appears between the mean probabilities associated with credibility ratings \textbf{1}, \textbf{2} and \textbf{3} whose directionality is judged as clearly \textit{positive},\footnote{i.e. “Confirmed”, “Probably True”, “Possibly True”.} and the mean probabilities associated with ratings \textbf{4} and \textbf{5} whose directionality is seen as clearly \textit{negative}.\footnote{i.e. “Doubtful”, “Improbable”.} Similarly, they observed a clear inflection between the mean probabilities of reliability ratings: \textbf{A}, \textbf{B} and \textbf{C} whose directionality is perceived as clearly \textit{positive}\footnote{i.e. “Completely Reliable”, “Usually Reliable”, “Fairly Reliable”.} form a separate group from ratings \textbf{D} and \textbf{E} whose directionality is seen as clearly \textit{negative}.\footnote{i.e. “Not Usually Reliable”, “Unreliable”.} In our proposal, we adopt the same clear-cut distinction between positive versus negative groupings of the intelligence ratings.
\medskip

Data collected by \cite{mandel2022meta} tend to show that intermediary areas are associated with rating “Cannot Be Judged”, both for credibility (rating \textbf{6}) and for reliability (rating \textbf{F}). Looking more closely at the inflection points of their data, the mean probabilities associated with ratings \textbf{6} and \textbf{F} fall exactly between the probabilities of positively directional ratings, on the one hand, and the probabilities of negatively directional ratings, on the other hand. This indicates that professionals seem to use ratings \textbf{6} and \textbf{F}, i.e. “Cannot Be Judged”, to evaluate neutral areas in case of credibility and reliability. In the present article, we do not discuss this identification between neutral areas and ratings “Cannot Be Judged”, but we also assume the existence of such neutral groupings in case of credibility and reliability. Table \ref{tab:groupingmatrix} show the result of grouping the 6$\times$6 dimensions of the alphanumeric matrix into 3$\times$3 based on \cite{mandel2022meta}.

\begin{table}[H]
\centering
  \small
{\renewcommand{\arraystretch}{1.4}
\begin{tabular} {|c|c|ccccccc|c|ccccc|} 

 \hline
 
\multicolumn{2}{|c|}{\cellcolor{kugray5}\textbf{Message}} & \multicolumn{13}{c|}{\textit{\textbf{Content Credibility}}}\\

    \cline{3-15}
    \multicolumn{2}{|c|}{\cellcolor{kugray5}\textbf{Scores}} &&\textbf{1}&&\textbf{2}&&\textbf{3}&&\textbf{6}&&\textbf{4}&&\textbf{5}&\\
    \hline
    \multirow{6}{*}{\begin{turn}{90}\textit{\textbf{Source Reliability}}\ \ \end{turn}}& \textbf{A}&&\textbf{A1}&&\textbf{A2}&&\textbf{A3}&&\textbf{A6}&&\textbf{A4}&&\textbf{A5}&\\
                  & \textbf{B}&&\textbf{B1}&&\textbf{B2}&&\textbf{B3}&&\textbf{B6}&&\textbf{B4}&&\textbf{B5}&\\
    
                       & \textbf{C}&&\textbf{C1}&&\textbf{C2}&&\textbf{C3}&&\textbf{C6}&&\textbf{C4}&&\textbf{C5}&\\

\cline{2-15}
       & \textbf{F}&&\textbf{F1}&&\textbf{F2}&&\textbf{F3}&&\textbf{F6}&&\textbf{F4}&&\textbf{F5}&\\

\cline{2-15}
                      & \textbf{D}&&\textbf{D1}&&\textbf{D2}&&\textbf{D3}&&\textbf{D6}&&\textbf{D4}&&\textbf{D5}&\\
    
                     & \textbf{E}&&\textbf{E1}&&\textbf{E2}&&\textbf{E3}&&\textbf{E6}&&\textbf{E4}&&\textbf{E5}&\\
    
       \cline{2-11}
                   
    \hline

\end{tabular}}
\caption{The Alphanumeric Matrix after Grouping Dimensions into 3$\times$3 categories.}
\label{tab:groupingmatrix}
\end{table}

\subsubsection{Facts and Interpretations are Not Distinguished De Facto}
\label{sssec:notdistin}





Given the results described above, \cite{Icard2023a} has argued that the \textit{facts versus interpretations recommendation} is not respected in pratice. Some explanation can be found in the empirical observations just reviewed: for officers to mark the distinction between facts and interpretations along the 6$\times$6 scale, they should first have a clear understanding of the 6 dimensions of credibility and 6 dimensions of reliability underlying the scale. However, experimental results show that officers are confused about the intrinsinc meanings of those dimensions and their distinct levels of evaluation. Being confused as such, it is difficult to see how officers could use the alphanumeric scale to tease apart facts from interpertations in intelligence. Another explanation is more fundamental. The alphanumeric scale being evaluative by nature, its role is to help officers provide evaluations of information. To this end, the scale relies on dimensions of credibility and reliability which are evaluative and non-descriptive or factual. Quite naturally, this makes those dimensions and the resulting scale rather unsuitable for providing a factual, that is to say more objective, insight into pieces of information. 


\medskip
Based on those two arguments, \citep{Icard2023a} proposed to extract the descriptive scale which remains in the background of the 3$\times$3 evaluative scale. This background scale relies on the descriptive dimensions of truth and honesty that credibility and reliability are intended to evaluate (see subsection \ref{ssec:alpha1}). According to the groupings observed by \cite{mandel2022meta} in the evaluative case, truth and honesty are also grouped into three categories to result into a 3$\times$3 descriptive matrix, or taxonomy, of intelligence messages. This taxonomy is presented in more details in subsection \ref{ssec:informalicard} and retrieved in section \ref{sec:classifying} as a result of the new evaluative procedure we present in section \ref{sec:new}. From now own, let us keep in mind that the goal of this descriptive taxonomy was to make officers meet the \textit{facts versus interpretations recommendation} by helping them provide a descriptive insight on intelligence messages in addition to their more personal evaluations.

\subsubsection{Credibility and Reliability are Not Independent De Jure} 
\label{sssec:notinde}



With the extant procedure, not only are officers unable to follow the \textit{facts versus interpretations recommendation} \textit{de facto}, but they are unable to respect the \textit{credibility versus reliability recommendation} \textit{de jure}. As we have seen, \cite{Baker&al1968} observed that the large majority of resultant scores fall on the diagonal continuum of the scale, that is on the line \textbf{A1}-\textbf{B2}-\textbf{C3}-\textbf{D4}-\textbf{E5}-\textbf{F6}. In fact, not only does this indicate that officers only use a subpart of the scale, it also shows that officers do not perceive credibility and reliability as independent dimensions but as strongly correlated. In 1975, this correlation observed by \cite{Baker&al1968} was replicated by \cite{Samet1975} who also conducted further analyses to determine which of the credibility and reliability dimensions was prevalent in resultant scores. 

\medskip
Samet's investigations followed two paths in particular. In one experiment, 37 intelligence officers were asked to express the conditional probability that a given report carried a specific \textit{credibility} (or \textit{reliability}) rating provided that the report \textit{already} carried a specific \textit{reliability} (or \textit{credibility}) rating. Results proved to be highly significant 73\% of the time since reponses provided by 27 officers showed a strong interaction between the probabilities the officers assigned to credibility and reliability ratings.\footnote{For exactly 10 officers, however, the two dimensions were treated independently. This observation, however, has not been included in Samet's analysis because officers have different conceptualizations of the ratings in this case. For these officers, the credibility of the content gives no indication of the reliability of the source who delivered that content, — which is counterintuitive according to the doctrinal descriptions (see Table \ref{tab:cred}). Conversely, the reliability of the source is not seen as the determinant parameter for estimating the credibility of the content she delivers, — which is also counterintuitive based on the doctrinal descriptions (see Table \ref{tab:rel}).}

\medskip

In another experiment, Samet aimed to estimate the relative influence of the individual credibility and reliability ratings in joint ratings. Setting aside ratings \textbf{6} and \textbf{F}, Samet asked the 37 intelligence officers to assign a probability degree to each of the 25 possible scores of the 5$\times$5 evaluative part the of the scale. They also asked them to assign a probability degree to each of the 5 credibility ratings and to each of the 5 reliability ratings, independently. Multiple linear regression reveals that for 35 of the 37 officers, credibility accounted for 76.6\% of resultant scores whereas reliability only accounted for 23.4\% of them \citep[see][]{Samet1975}. In other words, the credibility dimension was considered \textit{three times as important as} the reliability dimension by intelligence officers.\footnote{As a matter of fact, results from \cite{Baker&al1968} already gave a slight indication of this prevalence. Setting aside the diagonal continuum (which comprised 87\% of joint ratings), scores were mostly \textit{confined to the high end of the scale}, thus indicating a weak preference for the credibility dimension over the reliability dimension. In fact, 11\% of the scores fell (stricly) above the diagonal of the scale ($N$ = 75 out of 695) whereas only 2\% of the scores fell (strictly) below the diagonal ($N$ = 12 out of 695). But Samet's results gives a clear-cut insight into the dominance of the credibility dimension.} This dominance was also observed by \cite{Miron&al1978} in a study with 55 officers enrolled to rate 40 different messages in an intelligence course. They found that credibility accounted for 57\% of the quality of intelligence messages, even though other dimensions also played a role of lesser importance, such as \textit{relevance} (19\%) and \textit{directness} (6\%).
\medskip

Practically then, these results show that the evaluative procedure must be revised. Following recommendations from \cite{Samet1975} and \cite{Phelps&al1980}, intelligence evaluation should be made along a \textit{single credibility scale} on which reliability plays a a secondary role. In the next section, we propose a new procedure for intelligence evaluation which aims to keep the virtues of the existing scale but to remedy its shortcomings.

\section{Extant Proposals to Improve Information Evaluation}
\label{sec:issues}





















\subsection{Informal Proposals to Help Clarify the Scale Dimensions}

\label{ssec:informalicard}




Broadly speaking, the descriptive proposal made by \cite{Icard2023a} consists in isolating the more objective, that is to say more descriptive, background hidden within the extant alphanumeric procedure in order to propose a descriptive taxonomy of intelligence messages. Table \ref{tab:associate}, we put emphasis on the correspondence between the 3$\times$3 evaluative matrix obtained by grouping the credibility and reliability dimensions into three categories \citep[based on][]{teigen&brun1999directionality,mandel2022meta}, and the 3$\times$3 descriptive matrix it corresponds to when dimensions of truth and honesty are subjected to the same treatment.

\begin{table}[H]
\centering
  \small
{\renewcommand{\arraystretch}{1.4}
\begin{tabular} {|c|c|ccccccccccccc|} 

 \hline
 
\multicolumn{2}{|c|}{\cellcolor{kugray5}\textbf{Message}} & \multicolumn{13}{c|}{\textit{\textbf{Content Credibility}}}\\

    \cline{3-15}
    \multicolumn{2}{|c|}{\cellcolor{kugray5}\textbf{Scores}} &&\textbf{1}&&\textbf{2}&&\textbf{3}&|&\textbf{6}&|&\textbf{4}&&\textbf{5}&\\
    \hline
    \multirow{6}{*}{\begin{turn}{90}\textit{\textbf{Source Reliability}}\ \ \end{turn}}& \textbf{A}&&&&&&&&&&&&&\\
    
                  & \textbf{B}&&&&&&&&&&&&&\\

                       & \textbf{C}&&&&&&&&&&&&&\\

\cline{2-2}

       & \textbf{F}&&&&&&&&&&&&&\\

\cline{2-2}
                      & \textbf{D}&&&&&&&&&&&&&\\

                     & \textbf{E}&&&&&&&&&&&&&\\

    \hline

\end{tabular}}
\caption{The Correspondence between the 3$\times$3 Evaluative Matrix given in Table \ref{tab:groupingmatrix}, and the 3$\times$3 Descriptive Dimensions of Truth and Honesty isolated in \cite{Icard2023a}.}
\label{tab:associate}
\end{table}

\begin{tikzpicture}
\tikz[remember picture,overlay]{
  \draw[decorate,decoration={brace,mirror,amplitude=2pt,raise=2pt}] (5.22,6.2) --node[below=5pt] {True} (8.4,6.2);
}
\tikz[remember picture,overlay]{
  \draw[decorate,decoration={brace,mirror,amplitude=2pt,raise=2pt}] (8.6,6.2) --node[below=6pt] {\scalebox{0.9}{Indeterminate}} (9.6,6.2);
}
\tikz[remember picture,overlay]{
  \draw[decorate,decoration={brace,mirror,amplitude=2pt,raise=2pt}] (9.8,6.2) --node[below=5pt] {False} (12,6.2);
}
\tikz[remember picture,overlay]{
  \draw[decorate,decoration={brace,amplitude=2pt,raise=2pt}] (5.07,6.05) --node[right=5pt] {Honest} (5.07,4.2);
}
\tikz[remember picture,overlay]{
  \draw[decorate,decoration={brace,amplitude=2pt,raise=2pt}] (5.07,4.1) --node[right=5pt] {Imprecise} (5.07,3.5);
}
\tikz[remember picture,overlay]{
  \draw[decorate,decoration={brace,amplitude=2pt,raise=2pt}] (5.07,3.45) --node[right=5pt] {Dishonest} (5.07,2.2);
}
\end{tikzpicture}

From these two descriptive dimensions of truth and honesty underlying the extant evaluative scale, \cite{Icard2023a} has proposed a taxonomy of intelligence messages, based on adaptating the taxonomy of informational vagueness presented in \cite{Egre&Icard2018}.\footnote{See \cite{egre2018vague} for a detailed presentation of the different theories of linguistic vagueness and \cite{egre2020optimality} in particular on the optimal use of vague language in communication.} The intelligence taxonomy presented in \cite{Icard2023a} aimed to help officers respect the \textit{facts versus interpretations recommendation}. This taxonomy is based on two dimensions, that is to say\textit{“Truth of the Content”} vs. \textit{“Honesty of the Source”}, and adapts Mandel et al.'s observation at the descriptive level, thus distinguishing three levels of truth for message contents (\textit{“True”}, \textit{“False”}, \textit{“Indeterminate”}) and three levels of honesty for sources (\textit{“Honest”}, \textit{“Dishonest”}, \textit{“Imprecise”}). With regard to the truth dimension, a content is said to be \textit{true} if it corresponds to objective facts, \textit{false} if it does not and \textit{indeterminate} if the content is semantically vague. With regard to the honesty dimension, sources are said to be \textit{honest} in case they tell exactly what they believe to be true, \textit{dishonest} if they tell the opposite of what they believe and \textit{imprecise} if they are pragmatically vague. This 3$\times$3 taxonomy results in nine categories associated with more well-known informational labels from the litterature. 

\medskip

Table \ref{tab:icardtaxo} gives a complete overview of this descriptive taxonomy in which message types are based either on precise dimensions (\textit{true} versus \textit{false} contents, \textit{honest} versus \textit{dishonest} sources) and/or on vague dimensions (\textit{indeterminate} content, \textit{imprecise} source). Precise dimensions concern types $\textbf{\textit{\texttt{t}}}_{\textbf{\texttt{\textit{1}}}}$, $\textbf{\textit{\texttt{t}}}_{\textbf{\texttt{\textit{2}}}}$, $\textbf{\textit{\texttt{t}}}_{\textbf{\texttt{\textit{3}}}}$ and $\textbf{\textit{\texttt{t}}}_{\textbf{\texttt{\textit{4}}}}$ because in those cases, contents are semantically precise, being either true or false, while sources are pragmatically precise, being either honest or dishonest, with no extra possibilities. Vague dimensions concern types $\textbf{\textit{\texttt{t}}}_{\textbf{\texttt{\textit{5}}}}$, $\textbf{\textit{\texttt{t}}}_{\textbf{\texttt{\textit{6}}}}$, $\textbf{\textit{\texttt{t}}}_{\textbf{\texttt{\textit{7}}}}$, $\textbf{\textit{\texttt{t}}}_{\textbf{\texttt{\textit{8}}}}$ and $\textbf{\textit{\texttt{t}}}_{\textbf{\texttt{\textit{9}}}}$ because message contents can be semantically indeterminate, being borderline between truth and falsity (as in $\textbf{\textit{\texttt{t}}}_{\textbf{\texttt{\textit{5}}}}$ and $\textbf{\textit{\texttt{t}}}_{\textbf{\texttt{\textit{6}}}}$), or because sources can be pragmatically imprecise, being borderline between honesty and dishonesty (as in $\textbf{\textit{\texttt{t}}}_{\textbf{\texttt{\textit{7}}}}$ and $\textbf{\textit{\texttt{t}}}_{\textbf{\texttt{\textit{8}}}}$). Type $\textbf{\textit{\texttt{t}}}_{\textbf{\texttt{\textit{9}}}}$ is a mixed case in which both the content and source are vague. 
From now on, we consider that indeterminate contents and imprecise sources are only borderline but not clearly defective. By contrast, we consider false contents and dishonest sources to be clearly defective.

\medskip

\begin{table}[h!]
  \centering
\begin{tabular}{|c|c|c|c|c|}

 \hline
 
\multicolumn{2}{|c|}{\cellcolor{kugray5} \textbf{Message Type}} & \multicolumn{3}{c|}{\textit{\textbf{Truth of the Content}}\ \ \ \ \ }\\

    \cline{3-5}
    \multicolumn{2}{|c|}{\cellcolor{kugray5}\textit{\textbf{\texttt{t}}}} & \textit{True} & \textit{Indeterminate} & \textit{False} \\ 
    \hline
    &&&& \\ 
    \multirow{2}{*}{\begin{turn}{90} \textit{\textbf{Honesty of the Source}} \ \ \ \ \ \ \ \end{turn}}& &$\textbf{\textit{\texttt{t}}}_{\textbf{\texttt{\textit{1}}}}$&
    $\textbf{\textit{\texttt{t}}}_{\textbf{\texttt{\textit{5}}}}$&
    $\textbf{\textit{\texttt{t}}}_{\textbf{\texttt{\textit{2}}}}$\\
    
    &\textit{Honest}& = & = & = \\
    
   & &\textit{information}&\textit{error-avoidance}&\textit{misinformation} \\  
&&&& \\  
       \cline{2-5}
&&&& \\ 
& &$\textbf{\textit{\texttt{t}}}_{\textbf{\texttt{\textit{7}}}}$ & $\textbf{\textit{\texttt{t}}}_{\textbf{\texttt{\textit{9}}}}$& $\textbf{\textit{\texttt{t}}}_{\textbf{\texttt{\textit{8}}}}$\\
                        
 &\textit{Imprecise}& = & = & = \\
                    
   & &\textit{omission}&\textit{mixed}&\textit{dissimulation} \\  
&&&& \\  
       \cline{2-5}
&&&& \\ 
                                           & &$\textbf{\textit{\texttt{t}}}_{\textbf{\texttt{\textit{3}}}}$&$\textbf{\textit{\texttt{t}}}_{\textbf{\texttt{\textit{6}}}}$&$\textbf{\textit{\texttt{t}}}_{\textbf{\texttt{\textit{4}}}}$\\
                        
                        &\textit{Dishonest}& = & = & = \\
                    
   & &\textit{subjective lie}&\textit{half-truth}&\textit{objective lie} \\  
&&&& \\  
      
    \hline
   
  \end{tabular}
  \centering{\caption{The Descriptive Taxonomy of Intelligence Messages presented in \cite{Icard2023a}.}\label{tab:icardtaxo}}

\end{table}

The nine categories of the taxonomy can in fact be ranked \textit{qualitatively} from best to worse. Quite intuitively, the quality of a message improves when the credibility of its content increases and/or when the reliability of its source increases. By contrast, the quality of a message decreases when its credibility decreases and when its source becomes less reliable. Accordingly, we propose the following qualitative ranking (\textbf{QL}) of the nine message types of the descriptive taxonomy: 

\begin{center} (\textbf{QL}) \hspace{0.5cm}  
$\textbf{\textit{\texttt{t}}}_{\textbf{\texttt{\textit{1}}}} >_{ql} \textbf{\textit{\texttt{t}}}_{\textbf{\texttt{\textit{5}}}} >_{ql}  \textbf{\textit{\texttt{t}}}_{\textbf{\texttt{\textit{7}}}} >_{ql} \textbf{\textit{\texttt{t}}}_{\textbf{\texttt{\textit{2}}}} >_{ql} \textbf{\textit{\texttt{t}}}_{\textbf{\texttt{\textit{9}}}} >_{ql} 
\textbf{\textit{\texttt{t}}}_{\textbf{\texttt{\textit{3}}}} >_{ql} 
\textbf{\textit{\texttt{t}}}_{\textbf{\texttt{\textit{8}}}} >_{ql} 
\textbf{\textit{\texttt{t}}}_{\textbf{\texttt{\textit{6}}}} >_{ql} 
\textbf{\textit{\texttt{t}}}_{\textbf{\texttt{\textit{4}}}}$
\end{center}
where ``$>_{ql}$'' means that the informational type on the left of the inequality symbol is qualitatively better than the informational type on the right of the symbol.

\medskip
The qualitative ranking (\textbf{QL}) orders both precise and vague message types. Concerning precise types (i.e. $\textbf{\textit{\texttt{t}}}_{\textbf{\texttt{\textit{1}}}}$, $\textbf{\textit{\texttt{t}}}_{\textbf{\texttt{\textit{2}}}}$, $\textbf{\textit{\texttt{t}}}_{\textbf{\texttt{\textit{3}}}}$ and $\textbf{\textit{\texttt{t}}}_{\textbf{\texttt{\textit{4}}}}$), the relative ordering is $\textbf{\textit{\texttt{t}}}_{\textbf{\texttt{\textit{1}}}} >_{ql} ... >_{ql} \textbf{\textit{\texttt{t}}}_{\textbf{\texttt{\textit{2}}}} >_{ql} ... >_{ql} \textbf{\textit{\texttt{t}}}_{\textbf{\texttt{\textit{3}}}} >_{ql} ... >_{ql} \textbf{\textit{\texttt{t}}}_{\textbf{\texttt{\textit{4}}}}$ because, qualitatively, $\textbf{\textit{\texttt{t}}}_{\textbf{\texttt{\textit{1}}}}$ is the most valuable type (true content, honest source) of the entire ordering (\textbf{QL}) while $\textbf{\textit{\texttt{t}}}_{\textbf{\texttt{\textit{4}}}}$ is the least valuable type (false content, dishonest source) of the ordering. Types $\textbf{\textit{\texttt{t}}}_{\textbf{\texttt{\textit{2}}}}$ (false content, honest source) and $\textbf{\textit{\texttt{t}}}_{\textbf{\texttt{\textit{3}}}}$ (true content, dishonest source) are qualitatively intermediate because both types have at least one defective dimension: truth of the content in $\textbf{\textit{\texttt{t}}}_{\textbf{\texttt{\textit{2}}}}$, honesty of the source in $\textbf{\textit{\texttt{t}}}_{\textbf{\texttt{\textit{3}}}}$. But $\textbf{\textit{\texttt{t}}}_{\textbf{\texttt{\textit{2}}}}$ is qualitatively better than $\textbf{\textit{\texttt{t}}}_{\textbf{\texttt{\textit{3}}}}$ because the source being honest in $\textbf{\textit{\texttt{t}}}_{\textbf{\texttt{\textit{2}}}}$, she is also deceived by the content of her message: this is a case of honest mistake. By contrast, the content in $\textbf{\textit{\texttt{t}}}_{\textbf{\texttt{\textit{3}}}}$ is true but the source is dishonest, so her intentions regarding the officer are deceptive. For this reason, $\textbf{\textit{\texttt{t}}}_{\textbf{\texttt{\textit{3}}}}$ is lower than $\textbf{\textit{\texttt{t}}}_{\textbf{\texttt{\textit{2}}}}$ in the qualitative ordering of precise types. 

\medskip

Concerning vague types, (i.e. $\textbf{\textit{\texttt{t}}}_{\textbf{\texttt{\textit{5}}}}$, $\textbf{\textit{\texttt{t}}}_{\textbf{\texttt{\textit{6}}}}$, $\textbf{\textit{\texttt{t}}}_{\textbf{\texttt{\textit{7}}}}$, $\textbf{\textit{\texttt{t}}}_{\textbf{\texttt{\textit{8}}}}$ and $\textbf{\textit{\texttt{t}}}_{\textbf{\texttt{\textit{9}}}}$), the relative ordering is $... >_{ql} \textbf{\textit{\texttt{t}}}_{\textbf{\texttt{\textit{5}}}} >_{ql}  \textbf{\textit{\texttt{t}}}_{\textbf{\texttt{\textit{7}}}} >_{ql} ... >_{ql} \textbf{\textit{\texttt{t}}}_{\textbf{\texttt{\textit{9}}}} >_{ql} ... >_{ql}\textbf{\textit{\texttt{t}}}_{\textbf{\texttt{\textit{8}}}} >_{ql}  \textbf{\textit{\texttt{t}}}_{\textbf{\texttt{\textit{6}}}} >_{ql} ... $. Types $\textbf{\textit{\texttt{t}}}_{\textbf{\texttt{\textit{5}}}}$ and $\textbf{\textit{\texttt{t}}}_{\textbf{\texttt{\textit{7}}}}$ are higher in the ordering than $\textbf{\textit{\texttt{t}}}_{\textbf{\texttt{\textit{6}}}}$ and $\textbf{\textit{\texttt{t}}}_{\textbf{\texttt{\textit{8}}}}$. In $\textbf{\textit{\texttt{t}}}_{\textbf{\texttt{\textit{5}}}}$ and $\textbf{\textit{\texttt{t}}}_{\textbf{\texttt{\textit{7}}}}$, one dimension is qualitatively borderline without being clearly defective (indeterminate content for $\textbf{\textit{\texttt{t}}}_{\textbf{\texttt{\textit{5}}}}$, imprecise source for $\textbf{\textit{\texttt{t}}}_{\textbf{\texttt{\textit{7}}}}$), and the other dimension is clearly good in both cases (honest source for $\textbf{\textit{\texttt{t}}}_{\textbf{\texttt{\textit{5}}}}$, true content for $\textbf{\textit{\texttt{t}}}_{\textbf{\texttt{\textit{7}}}}$). In types $\textbf{\textit{\texttt{t}}}_{\textbf{\texttt{\textit{6}}}}$ and $\textbf{\textit{\texttt{t}}}_{\textbf{\texttt{\textit{8}}}}$, by contrast, not only is one dimension borderline (indeterminate content for $\textbf{\textit{\texttt{t}}}_{\textbf{\texttt{\textit{6}}}}$, imprecise source for $\textbf{\textit{\texttt{t}}}_{\textbf{\texttt{\textit{8}}}}$) but the other dimension is now clearly defective in both cases (dishonest source for $\textbf{\textit{\texttt{t}}}_{\textbf{\texttt{\textit{6}}}}$, false content for $\textbf{\textit{\texttt{t}}}_{\textbf{\texttt{\textit{8}}}}$). In type $\textbf{\textit{\texttt{t}}}_{\textbf{\texttt{\textit{9}}}}$ (indeterminate content, imprecise source), both dimensions are borderline but none of them is clearly defective. For this reason, $\textbf{\textit{\texttt{t}}}_{\textbf{\texttt{\textit{9}}}}$ is qualitatively intermediate between $\textbf{\textit{\texttt{t}}}_{\textbf{\texttt{\textit{5}}}}$ and $\textbf{\textit{\texttt{t}}}_{\textbf{\texttt{\textit{7}}}}$ in which only one dimension is borderline and types $\textbf{\textit{\texttt{t}}}_{\textbf{\texttt{\textit{6}}}}$ and $\textbf{\textit{\texttt{t}}}_{\textbf{\texttt{\textit{8}}}}$ in which one dimension is borderline while the other is clearly defective. That being said, $\textbf{\textit{\texttt{t}}}_{\textbf{\texttt{\textit{9}}}}$ is worst than $\textbf{\textit{\texttt{t}}}_{\textbf{\texttt{\textit{2}}}}$ because in $\textbf{\textit{\texttt{t}}}_{\textbf{\texttt{\textit{9}}}}$, both content and source are borderline while only the content is defective in $\textbf{\textit{\texttt{t}}}_{\textbf{\texttt{\textit{2}}}}$ but this is not intentional: the source is honest in $\textbf{\textit{\texttt{t}}}_{\textbf{\texttt{\textit{2}}}}$. By contrast, $\textbf{\textit{\texttt{t}}}_{\textbf{\texttt{\textit{9}}}}$ is better than $\textbf{\textit{\texttt{t}}}_{\textbf{\texttt{\textit{3}}}}$ because in $\textbf{\textit{\texttt{t}}}_{\textbf{\texttt{\textit{3}}}}$, the content is true but the source is clearly dishonest and deceptive.

\medskip

Type $\textbf{\textit{\texttt{t}}}_{\textbf{\texttt{\textit{5}}}}$ is qualitatively better than type $\textbf{\textit{\texttt{t}}}_{\textbf{\texttt{\textit{7}}}}$, i.e. $\textbf{\textit{\texttt{t}}}_{\textbf{\texttt{\textit{5}}}} >_{ql} \textbf{\textit{\texttt{t}}}_{\textbf{\texttt{\textit{7}}}}$. Both types rely only on one borderline dimension (indeterminate content for $\textbf{\textit{\texttt{t}}}_{\textbf{\texttt{\textit{5}}}}$, imprecise source for $\textbf{\textit{\texttt{t}}}_{\textbf{\texttt{\textit{7}}}}$). But in type $\textbf{\textit{\texttt{t}}}_{\textbf{\texttt{\textit{7}}}}$, imprecision is used to hide the truth intentionally (\textit{omission}) while in $\textbf{\textit{\texttt{t}}}_{\textbf{\texttt{\textit{5}}}}$, indeterminacy is used honestly in order to avoid communicating borderline information that could be false and misleading. Symmetrically, type $\textbf{\textit{\texttt{t}}}_{\textbf{\texttt{\textit{8}}}}$ is qualitatively better than type $\textbf{\textit{\texttt{t}}}_{\textbf{\texttt{\textit{6}}}}$, i.e. $\textbf{\textit{\texttt{t}}}_{\textbf{\texttt{\textit{8}}}} >_{ql}  \textbf{\textit{\texttt{t}}}_{\textbf{\texttt{\textit{6}}}}$. Now both types rely on a defective dimension (false content for $\textbf{\textit{\texttt{t}}}_{\textbf{\texttt{\textit{6}}}}$, dishonest source for $\textbf{\textit{\texttt{t}}}_{\textbf{\texttt{\textit{8}}}}$), while the other dimension is borderline (indeterminate content for $\textbf{\textit{\texttt{t}}}_{\textbf{\texttt{\textit{6}}}}$, imprecise source for $\textbf{\textit{\texttt{t}}}_{\textbf{\texttt{\textit{8}}}}$). But $\textbf{\textit{\texttt{t}}}_{\textbf{\texttt{\textit{6}}}}$ is qualitatively worse than $\textbf{\textit{\texttt{t}}}_{\textbf{\texttt{\textit{8}}}}$ since in $\textbf{\textit{\texttt{t}}}_{\textbf{\texttt{\textit{6}}}}$, the source is clearly dishonest and deceptive while in $\textbf{\textit{\texttt{t}}}_{\textbf{\texttt{\textit{6}}}}$, the content of the message is false but the source is not dishonest and deceptive, but only imprecise regarding the information she communicates.

\medskip
Since the 3$\times$3 taxonomy given in Table \ref{tab:icardtaxo} intends to capture the descriptive background of the 3$\times$3 evaluative scale (see Table \ref{tab:groupingmatrix}), we may wonder how those two scales correspond to one another. In other words, we might expect that an evaluative procedure defined in line with Table \ref{tab:groupingmatrix} helps retrieve the various informational types isolated in the taxonomy of Table \ref{tab:icardtaxo}. Before doing this, however, we will argue that the evaluative procedure itself, wether it is based on a 6$\times$6 matrix (according to the intelligence doctrine) or on a 3$\times$3 matrix (as observed in experimental results), should be revised first.

\subsection{Formal Proposals, Positive Aspects and Some Limits}
\label{ssec:extantformal}

Information evaluation has also aroused strong interest amongst computer scientists and logicians, especially those working on data security and intelligence processing. Drawing lines can be useful for clarity purposes, so we propose to review extant formal proposals by distinguishing \textit{quantitative} and \textit{qualitative approaches} to information evaluation.\footnote{Even though this distinction is artificial to some extent: quantitative approaches deal with qualitative notions (such as \textit{credibility}, \textit{reliability}, \textit{truth}, \textit{likelihood}, etc.), and most of the qualitative proposals have a quantitative flavour (through the elicitation of degrees for qualitative dimensions). However, this coarse delineation gives a clearer representation of the field of information evaluation \citep[see][for surveys]{Rogova2004,Capet&Delavallade2013,Revault&Lesot2017}.}

\subsubsection{Quantitative Proposals Based on Probabilistic Reasoning}





Quantitative approaches conceive information evaluation as a \textit{fusion issue} and define \textit{numerical} settings using Zadeh's possibility theory as starting point \citep{Zadeh1978,Dubois&Prade1990}, or probabilistic reasoning as in Bayesian analysis \citep{Bayes1763,Jeffreys1939,Pearl2014} or as in Dempster-Shafer’s theory of evidence \citep{Dempster1967,Shafer1976}. 

\medskip
Fusion operators based on possibility theory capture \textit{degrees of uncertainty} about epistemic and doxastic attitudes through \textit{possibility} and \textit{necessity measurements}. Different families of fusion operators have been proposed for aggregating imperfect and hetereogenous intelligence data in that case \citep[see][]{Lesot&al2011, Lesot&al2014}. These operators are defined as \textit{conjunctive} if the resultant score does not go beyond the minimum of initial values, \textit{disjunctive} if the result is greater or equal to the maximum of the arguments, based on a \textit{compromise} if the result is intermediary and, finally, \textit{variable} if the score alternates depending on the initial arguments. 

\medskip
Bayesian analysis has been used to help officers better appreciate the credibility of intelligence messages by using probability degrees instead of verbal quantifiers \citep[e.g.][]{Zlotnick1972,Fisk1972,Schweitzer1978,Schum1987,Barbieri2013,Blasch&al2013}. This approach was motivated by empirical data showing inconsistencies in individual and collective interpretations of the existing ratings. Probabilities are used to define \textit{prior} credibility ratings for message contents. Then, Bayesian rules are defined to \textit{update} these prior probabilities depending on incoming information one can use to cross-check the message content. But incoming information can also concern the reliability of the message source. In both cases, incoming information helps strike a balance on the prior probability distribution of ratings. 

\medskip
Fusion operators based on the Dempster-Shafer theory aim at helping officers compute the plausibility of some uncertain event, as well as the doxastic attitude to abide by, depending on data obtained from independent sources of various sensor types \citep[e.g.][]{Nimier2005,Cholvy2004info,Cholvy2010,Pichon&al2012}. Degrees of credibility are expressed by \textit{belief functions} rather than by Bayesian probability distributions. Probabilities encode evidence the officer has for particular messages. But these probabilities are assigned to sets of possible messages representing possible outcomes rather than to single and isolated messages.

\medskip
The aim of those quantitative approaches has been to mitigate the inconsistencies observed amongst officers using the 6$\times$6 scale, based on numerical probability degrees, Bayesian rules and fusion operators. However, as pointed out in subsection \ref{ssec:lackof}, substituing probabilities, or other kinds of numerical degrees, to the existing adverbial expressions used in ratings, as suggested by e.g. \cite{Kent1964}, is a solution also giving rise to between-raters inconsistencies \citep[see again][]{Wark1964,Johnson1973}. In addition to that, the interpretation of probabilities underlying the various quantitative approaches we presented, at least implicitly, is a subjective reading which conceives probabilities as degrees of belief.\footnote{See \cite{ramsey1926truth} and \cite{definetti2017theory} on the subjective interpretation of probability degrees.} As a consequence, it is difficult to see how these approaches can ensure that the \textit{facts versus interpretations recommendation} is respected, unless they also provide a more objective and descriptive perspective on information on top of the evaluation.     






\subsubsection{Qualitative Proposals Based on Non-Classical Logics}

Qualitative approaches to information evaluation are mainly \textit{symbolic} and based on \textit{non-classical logics}. Those qualitative approaches fall into two strands depending on whether they put an emphasis on the evaluation of message contents (by rating \textit{credibility}) or on the assessment of message sources (by rating \textit{reliability}). The formal proposal we make in this paper aims to take advantage of both strands.

\medskip
The first qualitative strand relies on \textit{many-valued logic} and consists in assessing the credibility of message contents by using more semantic values than the classical values of \textit{truth} (1) and \textit{falsity} (0). In many-valued logics, contents can receive extra discrete values, as in \textit{three-valued logics} or more, or values on a continuum from 0 to 1, as in \textit{fuzzy logics}. In both cases, values are interpreted epistemically: they intend to capture the agents' degrees of certainty and uncertainty on information quality. In those settings, the challenge is to define semantic clauses for the conjunction of message contents that may receive conflicting semantic values. Various combination rules have been proposed in this endeavour \citep[e.g.][]{Akdag&al1992,Seridi&Akdag2001,Revault&al2007,Revault&Lesot2015}. By adding up to an infinity of degrees to rate credibility, many-valued settings considerably increase the sensitivity of the evaluation process. But, in light of empirical results showing that officers tend to use less than six degrees of evaluation, those approaches may seem too expressive in relation to the officers' needs. 

\medskip
The second strand of qualitative approaches relies on \textit{modal logic}. Static epistemic operators have been defined to capture the \textit{beliefs}, \textit{desires} and \textit{intentions} of informational sources \citep[e.g.][]{Demolombe&Lorini2008,Herzig&al2010}. These operators can be combined to express \textit{profiles} of sources depending on their informational pedigree, that is on their disposition to deliver true messages (\textit{validity}) and to be maximally informative in doing so (\textit{completeness}). On a higher level, two types of unreliable sources are then characterized such as \textit{falsifiers} (who report \textit{false} information) and \textit{misinformers} (who report \textit{only} false information) \citep[see][]{Demolombe2004,Cholvy2013}. Based on these syntactic definitions for sources' profiles and types, axiomatic principles and inference rules are combined to assess new contents and sources through adequate derivations. In this perspective, information evaluation is conceived as a way of ascribing \textit{profiles} and \textit{types} to sources under investigation.  

\medskip 



Before presenting our own approach, we should mention that a mixed quantitative-qualitative approach has been proposed under the name of ``subjective logic'' \citep{josang2016subjective} with applications to intelligence analysis \citep{pope2005analysis,pope2010intelligence} and to information evaluation \citep{pope2006eval}. Combining Bayesian analysis, Dempster-Shafer theory and fuzzy logic, subjective logic is a multi-agent setting for modelling opinions through degrees of belief and disbelief (modulated by an uncertainty measure) and fusion operators for updating and revision opinions. The use of subjective logic to model information evaluation relies on updating the \textit{``collector’s assessment of source’s reliability''} (corresponding to reliability ratings from \textbf{A} to \textbf{F}, modelled as ``opinions'' combining degrees of belief, disbelief, uncertainty and base rate) with the \textit{``source’s assessment of the likelihood of information''} (corresponding to credibility ratings from \textbf{1} to \textbf{6}, modelled again as ``opinions'') \citep[see][p. 15]{pope2006eval}. The update of reliability ratings by credibility ratings are made through discounting and consensus operators which either decrease or increase the reliability of the source. 

\medskip

The subjective logic approach is very integrative and general. It relies on  extant qualitative and quantitative proposals and can be used to deal with multiple sources with varying degrees of reliability. In fact, subjective logic also extends to intelligence analysis helping weigh competing hypotheses. With respect to information evaluation, however, this framework is not perfectly in line with doctrinal recommendations and with empirical observations. First, as we pointed out before, the numerical encodings of belief, disbelief and uncertainty used in subjective logic can still give rise to inconsistencies within and between raters. Second, the subjective logical approach does not take into account that officers happen to use only a 3$\times$3 subpart of the 6$\times$6 Alphanumeric system by indicating a way to accomodate their approach to information evaluation with this empirical observation. Third, subjective logic being a combination of settings interpreting probabilities subjectively, the corresponding framework for scoring information is also subjective by nature and cannot provide the descriptive insight expected by the \textit{facts versus interpretations recommendation}. Finally, though this framework is consistent with the correlation observed between credibility and reliability \citep{Baker&al1968}, it runs counter the fact that reliability was in fact perceived by officers as an adjustement of credibility \citep{Samet1975,Miron&al1978,Phelps&al1980}, not the other way around.

\subsection{Our Formal Logic \Logic for Information Evaluation}
\label{ssec:proposal}






We propose a dynamic credibility revision logic called \Logic which combines doxastic modal logic with numerical belief revision to help improve information evaluation \cite{Aucher2004,vanDitmarsch2005,VanDitmarsch&Labuschagne2007}. In doing so, we aim to bridge the gap between qualitative and quantitative approaches on intelligence evaluation, and to connect descriptive and evaluative perspectives on this issue. 
\medskip




Our logical proposal must respect at least three requirements that we mentionned previously (see subsection \ref{ssec:strength/short}). First, logic \Logic should rely on clear and precise definitions of the main concepts of information evaluation: \textit{credibility}, \textit{reliability} but also the descriptive parameters of \textit{truth} and \textit{honesty} determining their evaluation. Second, the logic should help separate facts from interpretations, by contrast to the original intelligence procedure. Third, it should no longer separate credibility from reliability, but consider credibility as the first and foremost dimension of evaluation with reliability playing an adjustment dimension.    

\medskip

To limit the number of assumptions, however, we only take into account empirical findings on the credibility and reliability dimensions when defining the core syntactic and semantic features of \Logic. Credibility is expressed as the main syntactic and semantic feature of evaluation while the adjustement role of reliability is encoded semantically only. Also, we keep six degrees of evaluation both for credibility and for reliability, as in the original 6$\times$6 alphanumeric procedure. In other words, we do not group the 6$\times$6 dimensions of the scale into 3$\times$3 categories in our proposal, and in particular we do not assume that the ratings \textbf{6} and \textbf{F} correspond to intermediary evaluative dimensions in this case. Keeping our assumptions to a minimum, however, we show that \Logic can be adapted to the 3$\times$3 context easily. When doing so, we observe that \Logic perfectly agrees with the \textit{facts versus interpretations recommendation} since it can be used to retrieve the descriptive taxonomy of intelligence messages proposed in \cite{Icard2023a}.

\medskip
The remaining part of this article describe our evaluative procedure \Logic in details, presenting its syntax and semantic interpretation based on numerical belief revision theory (section \ref{sec:new}). That being done, we show that this new evaluative procedure integrates with the descriptive taxonomy proposed by \cite{Icard2023a} since, once used in its most general form, logic \Logic leads to retrieve the informational types described in this taxonomy (section \ref{sec:classifying}). In other words, \Logic sees information evaluation as a way of ascribing a descriptive category to the messages we are evaluating.

\section{A Dynamic Credibility Logic for Scoring Intelligence Messages}
\label{sec:new}

\subsection{The Interpreted Language \Logic}
\label{ssec:formal}

We define a dynamic doxastic language \Logic for assessing intelligence messages by considering the credibility of their contents and the reliability of their sources. Syntactically as well as semantically, our setting is inspired by previous works in qualitative belief revision that have a quantitative flavour\footnote{To use the words of \cite{Baltag&Smets2008logic} and of \cite{VanDitmarsch2008}.} \citep[see][]{Spohn1988,Aucher2004,Aucher2008,vanDitmarsch2005,VanDitmarsch2008,VanDitmarsch&Labuschagne2007}. 


\medskip
 Our language \Logic is a propositional syntax in which (conditional) credibility operators $\C^{n}$ and dynamic reliability operators $\R^\pm$ are logical primitives of the language. We start by defining the set \Form of all the \Logic-\textit{formulas} using the following Backus-Naur Form:  
\medskip

{\renewcommand{\arraystretch}{1.4}
\begin{tabular}{cccl}
$\langle$\textit{\textbf{Formulas}}$\rangle$ &  $\varphi,\psi$ &  :=  & $\top \ |\  \mathsf{intel} \ |\ \neg\varphi \ |\ \varphi \wedge \psi \ |\ \C^{n}\varphi\ |\ [\R^{\pm}\varphi]\psi$\\

$\langle$\textit{\textbf{Atoms}}$\rangle$ &  $\mathsf{intel}$ & := & $\mathsf{m} \mid \mathsf{e}$
\end{tabular}}

\medskip

where $n$ is an integer ranging from 1 to 6 that will be used to define six levels for credibility ratings: $n \in \{1,2,3,4,5,6\}$, and where $\pm$ is a parameter which varies from A to F with the aim to express reliability degrees: $\pm \in \{$A,B,C,D,E,F$\}$.\footnote{Note that when we present the 6$\times$6 alphanumeric scale (in Tables \ref{tab:scale}, \ref{tab:cred} and \ref{tab:rel}), and when we transform this 6$\times$6 scale into a 3$\times$3 scale (starting with Table \ref{tab:groupingmatrix}), we use bold fonts to refer to credibility ratings from \textbf{1} to \textbf{6} and to refer to reliability ratings from \textbf{A} to \textbf{F}. By contrast, in the syntax of \Logic, we use normal fonts to express credibility degrees from 1 to 6 and to express reliability updates from A to F. Based on semantic justifications, we will match credibility ratings \textbf{1} to \textbf{6} with syntactic notations 1 to 6 in subsection \ref{sssec:match-cred}, and reliability ratings \textbf{A} to \textbf{F} with syntactic notations A to F in subsection \ref{sssec:match-rel}.} The intelligence officer in charge of information evaluation is assumed but not expressed explicitly in \Logic. 

\medskip
In the syntax of \Logic, atomic formulas, abbreviated by $\mathsf{intel}$, can be of two types: $\mathsf{m}$ when $\mathsf{intel}$ is the propositional content of some informational message, and $\mathsf{e}$ when $\mathsf{intel}$ is a piece of evidence available for cross-checking messages. We will see later how atomic formulas $\mathsf{m}$ and $\mathsf{e}$ interact during information evaluation. Here we simply define $\mathsf{M}$ as the set of all atomic informational messages $\mathsf{m}$, $\mathsf{E}$ as the set of all atomic pieces of evidence $\mathsf{e}$ (for cross-checking informational messages $\mathsf{m}$) and $\mathsf{At}$ as the set of all atomic formulas \textit{stricto sensu}, so: $\mathsf{At} = \mathsf{M} \cup \mathsf{E}$. Complex formulas of \Logic are written $\varphi,\psi, \phi$, etc. Setting aside $\top$ and atomic formulas $\mathsf{intel}$, complex formulas can be the negation $\neg\varphi$ of an extant formula $\varphi$, the conjunction $\varphi \wedge \psi$ of two formulas $\varphi$ and $\psi$, from which additional propositional connectives ($\bot$,$\vee$,$\rightarrow$,$\leftrightarrow$) can be defined as usual. But more specifically, complex formulas can also be modal expressions consisting of a static operator, in case of $\C^{n}\varphi$, or of a dynamic operator, in case of $[\R^{\pm}\varphi]\psi$. 

\medskip
Before we go more into details, note that, intuitively, modal formulas $\C^{n}\varphi$ and $[\R^{\pm}\varphi]\psi$ should be interpreted along an epistemic reading. The intuitive interpretation of the (set of) conditional operators $\C^{n}\varphi$ is that the ``\textit{officer ascribes a degree of credibility n to formula $\varphi$}$"$. More precisely, $\C^{n}\varphi$ means that the officer ascribes a degree \textit{n} to formula $\varphi$ \textit{based on some piece(s) of evidence $\mathsf{e}\in$\Logic she detains in the context} (in a sense we will make precise). The intended reading of the (set of) dynamic operators $[\R^\pm\varphi]\psi$ is the following: ``\textit{After the officer performs a reliability update of type $\pm$ with formula $\varphi$, $\psi$ is the case}$"$.  


\medskip
We start out by giving a general semantic interpretation to \Logic in terms of doxastic plausibility models \citep{Vanbenthem2007,Baltag&Smets2006,Baltag&Smets2008,Vanbenthem&Smets2015}. Those structures will provide a starting point to express credibility rankings and updates of those rankings based on the reliability of the source.

\medskip
A plausibility model for \Logic is a relational structure $\mathds{S} = \ \langle \ \mathcal{S}, \leqslant,\ \parallel .\parallel\ \rangle$ such that:

\begin{itemize}
\item[\ding{118}] $\mathcal{S}$ is a finite non-empty set of ``\textit{possible states}$"$ \textit{\textbf{s}} (or ``\textit{worlds}$"$).
\smallskip
\item[\ding{118}] $\leqslant$ is a ``\textit{plausibility relation}'' such that $\leqslant\subseteq \mathcal{S}$\ $\times$\ $\mathcal{S}$. This relation $\leqslant$ is a well-preorder, that is to say a reflexive and transitive relation (\textit{preorder}) such that every non-empty subset of $\mathcal{S}$ has least elements (\textit{well-}preorder).\footnote{Well-foundedness is crucial here since it ensures that non-trivial formulas $\varphi$ are always conditionally believed on some formula $\psi$ \citep[see][for details]{VanDitmarsch2008,Baltag&Smets2006,Baltag&Smets2008}.}   

\smallskip
\item[\ding{118}] $\parallel .\parallel$:\ $\mathsf{At}\ \rightarrow\wp(\mathcal{S})$ is a standard ``\textit{valuation map}$"$ where $\mathsf{At}$ is the set of all atomic formulas $\mathsf{intel}$, and $\wp(\mathcal{S})$ is the set of all subsets of $\mathcal{S}$. 
\smallskip
\end{itemize}

 The preorder should be read as a (\textit{prior}) \textit{plausibility order}: when $\textit{\textbf{s}} \ \leqslant \ \textit{\textbf{t}}$ (for all $\textit{\textbf{s}},\textit{\textbf{t}}\in\mathcal{S}$), the intelligence officer considers that state \textit{\textbf{t}} is ``\textit{at least as plausible as}$"$ state \textit{\textbf{s}} in model $\mathds{S}$. We now provide semantic clauses to interpret the non-epistemic formulas of \Logic. The clauses for the credibility and reliability operators will be defined after having introduced numerical degrees in the current setting.  

\begin{itemize}
\item[•]$\mathds{S}, \textit{\textbf{s}} \models \top$ \textit{Always}. 
\item[•]$\mathds{S}, \textit{\textbf{s}} \models \mathsf{intel}$\ \ \textit{iff} \ \textit{\textbf{s}}$\in \parallel \mathsf{intel}\parallel$.
\item[•]$\mathds{S}, \textit{\textbf{s}} \models \neg \varphi\ \ \textit{iff}\ \ \mathds{S}, \textit{\textbf{s}} \not\models \varphi.$
\item[•]$\mathds{S}, \textit{\textbf{s}} \models\varphi\wedge \psi \ \textit{iff}\ \ \mathds{S}, \textit{\textbf{s}} \models \varphi \ $and$ \  \mathds{S}, \textit{\textbf{s}} \models \psi.$
\end{itemize}

 Note that according to the intelligence doctrine, pieces of evidence $\mathsf{e}$ are \textit{preconditions} for evaluating some information $\varphi$, and some atomic message $\mathsf{m}$ in particular. In that sense, evidence $\mathsf{e}$ can be either \textit{consistent} or \textit{inconsistent} with $\mathsf{m}$. Sometimes also, pieces of evidence $\mathsf{e}$ can be all false and cannot stand as preconditions for evaluating the credibility of $\mathsf{m}$. Semantically, we propose to capture these notions of \textit{consistency}, \textit{inconsistency} and \textit{failure} of adequate precondition by using the notion of \textit{conditional plausibility}. The more plausible a message $\mathsf{m}$ is relatively to some contextual evidence $\mathsf{e}$, the more evidence $\mathsf{e}$ can be seen as \textit{consistent} with $\mathsf{m}$ being true. On the contrary, the more the negation of message $\mathsf{m}$, namely $\neg\mathsf{m}$, is plausible based on some evidence $\mathsf{e}$, the more $\mathsf{e}$ can be seen as \textit{inconsistent} with $\mathsf{m}$ being true. When all pieces $\mathsf{e}$ are false, precondition fails to adjudicate on the credibility of $\mathsf{m}$

\subsection{Expressing Numerical Degrees to Rate Credibility}
\label{ssec:cred}
\subsubsection{Deriving Degrees from a Qualitative Ranking}

 Before matching credibility degrees with credibility ratings, we define $\parallel\mathsf{e}\parallel$ as the set of states of domain $\mathcal{S}$ that satisfy some piece of evidence $\mathsf{e}\in \mathsf{E}$, i.e. $\parallel\mathsf{e}\parallel := \{\textit{\textbf{u}}\  \in\mathcal{S}\ |\ \mathds{S}, \textit{\textbf{u}} \models \mathsf{e}\}$ for some $\mathsf{e}\in \mathsf{E}$. From the order $\leqslant$ and set $\parallel\mathsf{e}\parallel$, we can define 6 degrees of \textit{credibility strength} based on evidence $\mathsf{e}$. We write $\textbf{\textit{degree}}^{i}(\leqslant,\parallel\mathsf{e}\parallel)$ the set of all states of degree $i$ that also satisfy some piece of evidence $\mathsf{e}$. Since $\leqslant$ is a well-preorder, every non-empty subset of $\leqslant$ has maximal elements. Then, we can derive various degrees of credibility strength relative to evidence $\mathsf{e}$ by restricting further and further the set of $\leqslant$-maximal states that are $\mathsf{e}$-consistent.

\medskip
 Up to degree $n = 5$, sets $\textbf{\textit{degree}}^{n}(\leqslant,\parallel\mathsf{e}\parallel)$ are defined by induction as follows: 
\medskip

\begin{tabular}{llllllcll}

&&&& & $\textbf{\textit{degree}}^{1}(\leqslant,\parallel\mathsf{e}\parallel)$ & = & $\textit{Max}^{\parallel\mathsf{e}\parallel}_{\leqslant}$ & \\

&&& &&&& &\\
&&&& & $\textbf{\textit{degree}}^{n}(\leqslant,\parallel\mathsf{e}\parallel)$ & = 
& $Max^{\parallel\mathsf{e}\parallel\backslash\bigcup\limits_{i < n}\textbf{\textit{degree}}^{i}(\leqslant,\parallel\mathsf{e}\parallel)}_{\leqslant}$ & \textit{for some $\mathsf{e} \in \mathcal{F}_{(\textit{\texttt{intel}})}$} \\

\end{tabular}

\medskip
 For degree $\textit{n} = 6$, we define $\textbf{\textit{degree}}^{6}(\leqslant,\parallel\neg\mathsf{e}\parallel)$ by the clause:

\medskip

\begin{tabular}{llllllcll}

&&&& & $\textbf{\textit{degree}}^{6}(\leqslant,\parallel\neg\mathsf{e}\parallel)$ & = & $Max^{\parallel\neg\mathsf{e}\parallel}_{\leqslant}$ & \textit{for all $\mathsf{e} \in \mathcal{F}_{(\textit{\texttt{intel}})}$.}

\end{tabular}
\medskip

Here $\textit{Max}^{\parallel\mathsf{e}\parallel}_{\leqslant}$ is the set of states that satisfy $\mathsf{e}$ and are maximal for the ordering $\leqslant$ in model $\mathds{S}$. In other words, the officers considers those states to be the most plausible $\mathsf{e}$-states of the ordering. Given that $\parallel\mathsf{e}\parallel$ is the set of states of the ordering that satisfy evidence $\mathsf{e}$, we have $\textit{Max}^{\parallel\mathsf{e}\parallel}_{\leqslant} = \{$\textit{\textbf{u}} $\in\parallel\mathsf{e}\parallel\ |\ \forall$\textit{\textbf{v}}$\in\mathcal{S}$ \textit{\textbf{v}} $\leqslant$ \textit{\textbf{u}}$\}$. Contrariwise, $\textit{Max}^{\parallel\neg\mathsf{e}\parallel}_{\leqslant}$ is the set of the most plausible $\neg\mathsf{e}$-states of the ordering.

\medskip
 Intuitively, $\textbf{\textit{degree}}^{1}(\leqslant,\parallel\mathsf{e}\parallel)$ is the set of most plausible states that also satisfy \textit{some piece of evidence} $\mathsf{e} \in \mathsf{E}$: $\textit{Max}^{\parallel\mathsf{e}\parallel}_{\leqslant}$ for some $\mathsf{e} \in \mathsf{E}$. Then, the sets from $\textbf{\textit{degree}}^{2}(\leqslant,\ \parallel\mathsf{e}\parallel)$ to $\textbf{\textit{degree}}^{5}(\leqslant,\parallel\mathsf{e}\parallel)$ are obtained by successively \textit{removing} $\mathsf{e}$-states from the top of the plausibility ordering. 
In that sense, $\textbf{\textit{degree}}^{2}(\leqslant,\parallel\mathsf{e}\parallel)$ is the set of \textit{most plausible states} that are \textit{not the most plausible $\mathsf{e}$-states of the entire ordering}, namely that are not in $\textbf{\textit{degree}}^{1}(\leqslant,\parallel\mathsf{e}\parallel)$. Set $\textbf{\textit{degree}}^{3}(\leqslant,\parallel\mathsf{e}\parallel)$ is the set of most plausible $\mathsf{e}$-states that are \textit{neither} in $\textbf{\textit{degree}}^{1}(\leqslant,\parallel\mathsf{e}\parallel)$ \textit{nor} in $\textbf{\textit{degree}}^{2}(\leqslant,\parallel\mathsf{e}\parallel)$, \textit{and so on}. But $\textbf{\textit{degree}}^{6}(\leqslant,\parallel\neg\mathsf{e}\parallel)$ is defined differently: $\textbf{\textit{degree}}^{6}(\leqslant,\parallel\neg\mathsf{e}\parallel)$ is the set of the most plausible states of the entire ordering that do not satisfy \textit{piece of evidence} $\mathsf{e} \in \mathsf{E}$, namely $\textit{Max}^{\parallel\neg\mathsf{e}\parallel}_{\leqslant}$ for all $\mathsf{e} \in \mathsf{E}$.

\medskip
 Up to \textit{n} = 5, formula(s) $\mathsf{e}$ can be seen as an \textit{ordering source} in the sense of Kratzer' \citep[see][]{Kratzer1981,Kratzer1991,Lassiter2017}. Definitions of sets $\textbf{\textit{degree}}^{1}(\leqslant,\parallel\mathsf{e}\parallel)$ to $\textbf{\textit{degree}}^{5}(\leqslant,\parallel\mathsf{e}\parallel)$ are based on whether states that belong to them satisfy \textit{at least} one piece of evidence $\mathsf{e} \in \mathsf{E}$. But since $\textbf{\textit{degree}}^{1}(\leqslant,\parallel\mathsf{e}\parallel)$ to $\textbf{\textit{degree}}^{5}(\leqslant,\parallel\mathsf{e}\parallel)$ are obtained by successively removing $\mathsf{e}$-states from the top of the plausibility ordering, states are finally ordered depending on which formula(s) $\mathsf{e}$ they do satisfy or not. In that sense, formulas $\mathsf{e}$ induce a plausibility ordering over the distinct possible states depending on whether some evidence $\mathsf{e}$ turn out to be \textit{true} or \textit{false} at these states. This notion of \textit{ordering source} will become clearer in the evaluative case we present in subsection \ref{sssec:case1}.        

\medskip
 From those six sets of credibility strength, six degrees of conditional credibility can be defined for the intelligence officer. Up to degree \textit{n} = 5, the officer's conditional credibility of degree \textit{n} in formula $\varphi$ is given by the clause:

\begin{itemize}
\item[•]$\mathds{S}, \textit{\textbf{s}} \models \C^{n}\varphi$ \textit{iff} for all $\textit{\textbf{t}}\in \textbf{\textit{degree}}^{n}(\leqslant,\parallel\mathsf{e}\parallel)$: $\mathds{S}, \textit{\textbf{t}} \models \varphi$ for some $\mathsf{e} \in \mathsf{E}$.
\end{itemize}

 This semantic clause means that formula $\varphi$ is judged as credible by the officer at a degree \textit{n} $\leq$ 5 in state \textit{\textbf{s}} (of model $\mathds{S}$) \textit{if and only if} formula $\varphi$ is true in the most plausible states of degree \textit{n} that also satisfy some other piece of evidence $\mathsf{e}$. In other words, \textit{some precondition} $\mathsf{e}$ exists for evaluating the credibility of $\varphi$.

\medskip
 For degree \textit{n} = 6, conditional credibility is defined by:
\begin{itemize}
\item[•]$\mathds{S}, \textit{\textbf{s}} \models \C^{6}\varphi$ \textit{iff} for all $\textit{\textbf{t}}\in \textbf{\textit{degree}}^{6}(\leqslant,\parallel\neg\mathsf{e}\parallel)$: $\mathds{S}, \textit{\textbf{t}} \models \varphi$ for all $\mathsf{e} \in \mathsf{E}$.
\end{itemize}

\medskip

 The semantic clause means that formula $\varphi$ is judged as being credible to degree 6 by the officer at state \textit{\textbf{s}} (in model $\mathds{S}$) \textit{if and only if} formula $\varphi$ is true in the most plausible states of the ordering where no other piece of evidence $\mathsf{e}$ (distinct from $\varphi$ itself) also turns out to be true. In that case, \textit{precondition} for making an evaluation of $\varphi$ is lacking since all the pieces of evidence $\mathsf{e}$ are false.

\medskip
  For clarity's sake, let us rewrite the sets $\textbf{\textit{degree}}^{1}(\leqslant,\parallel\neg\mathsf{e}\parallel)$ to $\textbf{\textit{degree}}^{5}(\leqslant,\parallel\neg\mathsf{e}\parallel)$, as well as set $\textbf{\textit{degree}}^{6}(\leqslant,\parallel\neg\mathsf{e}\parallel)$, in a simpler way: $\textbf{\textit{degree}}^{i}$ for all $\textit{i} \in \{1,2,3,4,5,6\}$. From the sets $\textbf{\textit{degree}}^{i}$, we can define the degree function $\textit{\textbf{dg}}:\ \mathcal{S}\ \rightarrow\{1,2,3,4,5,6\}$ such that: $\textit{\textbf{dg}}(\textit{\textbf{s}}) = i \ \textit{iff}\ \textit{\textbf{s}}\in\textbf{\textit{degree}}^{i}$. This means that the degree of state \textit{\textbf{s}} is equal to \textit{i} if and only if state \textit{\textbf{s}} belongs to the set of states of degree \textit{i}. Accordingly, the semantic clause for operators $\C^{n}\varphi$ can be rewritten in the following way:       

\begin{itemize}
\item[•]$\mathds{S}, \textit{\textbf{s}} \models \C^{n}\varphi$ \textit{iff} for all \textit{\textbf{t}} such that $\textit{\textbf{dg}}(\textit{\textbf{t}}) = \textit{n}: \mathds{S}, \textit{\textbf{t}} \models \varphi$ with $\textit{n} \in \{1,2,3,4,5,6\}$.
\end{itemize}

\smallskip
 This clause intuitively means that formula $\varphi$ is judged as being credible by the officer to a degree \textit{n} $\leq$ 6 in state \textit{\textbf{s}} of model $\mathds{S}$ \textit{if and only if} formula $\varphi$ happens to be true in the states of degree \textit{n} of model $\mathds{S}$.  

\medskip
 Now we propose to match the various credibility operators with the doctrinal credibility ratings ranging from \textbf{1} to \textbf{6}. First note that the doctrinal ratings are \textit{not strict} but \textit{all conditional}, — at least when the evaluation is possible. According to Table \ref{tab:cred}, ascribing a level of credibility to a message $\mathsf{m}$ consists in determining the \textit{conditional credibility} of message $\mathsf{m}$ based on a set of \textit{contextual evidence} $\mathsf{e}$ the officer has for, or against, $\mathsf{m}$. In other words, we can capture doctrinal credibility ratings by distinguishing sets of consistent evidence which have different plausibility strength. Let us proceed step-by-step. 

\subsubsection{Matching Credibility Degrees with Credibility Ratings}
\label{sssec:match-cred}
 According to doctrinal descriptions (see Table \ref{tab:cred}), a message $\mathsf{m}$ is classified as ``Confir\-med$"$ and rated \textbf{1} if the (content of the) message is \textit{confirmed by independent sources} and \textit{consistent with other information on the subject}. Being \textit{confirmed}, the message is expected to reach the maximum credibility score. In such a case, $\mathsf{m}$ is judged as credible for being true in the \textit{best} set of plausible states satisfying some relevant evidence $\mathsf{e}$ consistent with it: $\textit{Max}^{\parallel\mathsf{e}\parallel}_{\leqslant}$ \textit{for some evidence $\mathsf{e} \in \mathsf{E}$}. Accordingly, we match this top credibility rating with conditional credibility $\C^{1}$ from language \Logic. 

\medskip
 By contrast, a message is classified as ``Probably True$"$ and rated \textbf{2} if it is \textit{not confirmed} but \textit{consistent with other information on the subject}. Semantically, this rating is weaker than rating \textbf{1} since, when rated \textbf{2}, a message is consistent with \textit{most} but \textit{not all} the highly plausible evidence the officer has. This can be understood as follows: the message $\mathsf{m}$ is true in the \textit{second} set of most plausible states satisfying some evidence $\mathsf{e}$ consistent with it. So the best set is no longer $\textit{Max}^{\parallel\mathsf{e}\parallel}_{\leqslant}$ but a slight restriction of it: $\textit{Max}^{\parallel\mathsf{e}\parallel\backslash\textbf{\textit{degree}}^{1}}_{\leqslant}$. We propose to match this credibility rating with conditional credibility operator $\C^{2}\mathsf{m}$.

\medskip
 In the same vein, a content is classified as ``Possibly True$"$ and rated \textbf{3} if it is \textit{not confirmed} and only \textit{consistent with some other information on the subject}. Now the set of consistent evidence is even more tenuous than for rating \textbf{2}: the officer judges that \textit{some} but not \textit{most} of the states that are consistent with $\mathsf{m}$ are highly plausible. Now, we propose to match credibility rating \textbf{3} with degree $\C^{3}$. In that case, the set of most plausible states is the \textit{third best} set overall: $\textit{Max}^{\parallel\mathsf{e}\parallel\backslash\{\textbf{\textit{degree}}^{1}\cup\textbf{\textit{degree}}^{2}\}}_{\leqslant}$. Accordingly, rating \textbf{3} is matched with conditional credibility $\C^{3}\mathsf{m}$.

\medskip
 The description provided with rating ``Doubtfully True$"$ is more difficult to interpret: a message $\mathsf{m}$ is rated \textbf{4} if it is \textit{not confirmed}, \textit{possible} but not conclusive since \textit{no other information on the subject} is available. However, this description cannot be taken at face value because of two difficulties. First, if a message was classified as ``Doubtfully True$"$ because no other information on the subject was available, rating \textbf{4} could not be distinguished from rating \textbf{6}. Second, words used for the label ``Doubtfully True$"$ conflict with the description itself: \textit{doubtfully} suggests that the truth status of the message is \textit{uncertain}. 

\medskip
As a matter of fact, the label ``Doubtfully True$"$ indicates that there is \textit{doubt} concerning the credibility of the message $\mathsf{m}$. In this case, $\mathsf{m}$ is only \textit{moderately plausible} based on the contextual evidence the officer has. Accordingly, the set of states in which $\mathsf{m}$ is true is weaker than before in terms of plausibility strength. So we can assume that the message $\mathsf{m}$ is rated \textbf{4} when $\mathsf{m}$ turns out to be true in the \textit{fourth best} plausible states satisfying some evidence $\mathsf{e}$ consistent with it: $\textit{Max}^{\parallel\mathsf{e}\parallel\backslash\{\textbf{\textit{degree}}^{1}\cup\textbf{\textit{degree}}^{2}\cup\textbf{\textit{degree}}^{3}\}}_{\leqslant}$. The corresponding credibility operator for rating \textbf{4} is then $\C^{4}\mathsf{m}$.     

\medskip
 Concerning rating ``Improbable$"$, the doctrine tells that a message is classified \textbf{5} if it is \textit{not confirmed} as well as \textit{contradicted by other information on the subject}. In such a case, $\mathsf{m}$ is very implausible based on the contextual evidence the officer has. Intuitively then, $\mathsf{m}$ is true only in the \textit{fifth} set of most plausible states that are also the least plausible states of the ordering: $\textit{Max}^{\parallel\mathsf{e}\parallel\backslash\{\textbf{\textit{degree}}^{1}\cup\textbf{\textit{degree}}^{2}\cup\textbf{\textit{degree}}^{3}\cup\textbf{\textit{degree}}^{4}\}}_{\leqslant}$. Quite naturally, the credibility operator for rating \textbf{5} is $\C^{5}\mathsf{m}$. 

\medskip

 Finally, rating \textbf{6} is ascribed when the credibility of message $\mathsf{m}$ ``Cannot Be Judged$"$: \textit{no basis exists for evaluating the validity of} $\mathsf{m}$ in this case. That is: there is no evidence $\mathsf{e}$, consistent or inconsistent with $\mathsf{m}$, on which the officer can condition his credibility on $\mathsf{m}$. we propose to match rating \textbf{6} with credibility operator $\C^{6}\mathsf{m}$ for this reason. The clauses we have given for expressing credibility ratings are summed up in Table \ref{tab:credsum}. 
 \medskip

\begin{table}[h]
  \centering
  
    \begin{tabular}{|c|c||c|}
    \hline
   \rowcolor{kugray5} & & \\
    \rowcolor{kugray5}\textbf{Rating} & \textbf{Label} & \textbf{Corresponding Set of States}\\
   \rowcolor{kugray5} & & \\
    \hline
    & & \\
     $\C^{1}\mathsf{m}$  & \small{Confirmed} & \small{$\textit{Max}^{\parallel\mathsf{e}\parallel}_{\leqslant}$} \textit{for some evidence $\mathsf{e} \in \mathsf{E}$} \\

        & &\\\hline
    & & \\
     $\C^{2}\mathsf{m}$ & \small{Probably True} & \small{$\textit{Max}^{\parallel\mathsf{e}\parallel}_{\leqslant}\backslash\{\textbf{\textit{degree}}^{1}\}$ (\textit{id.})} \\
     & & \\\hline
    & & \\
      $\C^{3}\mathsf{m}$  & \small{Possibly True} & \small{$\textit{Max}^{\parallel\mathsf{e}\parallel}_{\leqslant}\backslash\{\textbf{\textit{degree}}^{1}\cup\textbf{\textit{degree}}^{2}\}$ (\textit{id.})} \\
     & & \\\hline
    & & \\
      $\C^{4}\mathsf{m}$ & \small{Doubtfully True} & \small{$\textit{Max}^{\parallel\mathsf{e}\parallel}_{\leqslant}\backslash\{\textbf{\textit{degree}}^{1}\cup\textbf{\textit{degree}}^{2}\cup\textbf{\textit{degree}}^{3}\}$ (\textit{id.})} \\
     & & \\\hline
    & & \\
      $\C^{5}\mathsf{m}$ & \small{Improbable} & \small{$\textit{Max}^{\parallel\mathsf{e}\parallel}_{\leqslant}\backslash\{\textbf{\textit{degree}}^{1}\cup\textbf{\textit{degree}}^{2}
      \cup\textbf{\textit{degree}}^{3}\cup\textbf{\textit{degree}}^{4}\}$ (\textit{id.})}\\
    & & \\\hline
    & & \\
      $\C^{6}\mathsf{m}$ & \small{Cannot Be Judged} & \small{$\parallel\neg\mathsf{e}\parallel$ \textit{for all evidence $\mathsf{e} \in \mathsf{E}$}} \\

    & & \\\hline

    \end{tabular}%
\caption{Semantic Conditions for the Credibility Ratings of a Given Message $\mathsf{m}$.}
\label{tab:credsum}
\end{table}%

Note that for credibility ratings $\C^{1}\mathsf{m}$ to $\C^{5}\mathsf{m}$, the definitions given in Table \ref{tab:credsum} are parameterized by the existence of some piece of evidence $\mathsf{e}$ for cross-checking message $\mathsf{m}$. This condition is expressed by the existential clause ``\textit{for some evidence $\mathsf{e} \in \mathsf{E}$}'' in case of $\C^{1}\mathsf{m}$. We do not repeat this clause, which is strictly the same, for $\C^{2}\mathsf{m}$ to $\C^{5}\mathsf{m}$. By contrast, for credibility rating $\C^{6}\mathsf{m}$, the definition is parameterized by the absence of any piece of evidence $\mathsf{e}$ for evaluating $\mathsf{m}$, so the condition is now quantified universally: ``\textit{for all evidence $\mathsf{e} \in \mathsf{E}$}''. More accurately, all the pieces of evidence that could be available for evaluating $\mathsf{m}$ turn out to be false in the context at stake.

\subsubsection{An Example with Nuclear Submarines}
\label{sssec:case1}

Let us model a case of information evaluation in logic \Logic. Graphically, to represent the fact that the state \textit{\textbf{t}} is \textit{at least as plausible as} the state \textit{\textbf{s}} for the officer (\textit{\textbf{s}} $\leqslant$ \textit{\textbf{t}}), we draw a right arrow ``$\rightarrow$''  from state \textit{\textbf{s}} to state \textit{\textbf{t}}: \textit{\textbf{s}}$\rightarrow$\textit{\textbf{t}}. When states \textit{\textbf{s}} and \textit{\textbf{t}} are \textit{equally plausible} for the officer (\textit{\textbf{s}} $\leqslant$ \textit{\textbf{t}} and \textit{\textbf{t}} $\leqslant$ \textit{\textbf{s}}), we draw a left-right arrow ``$\leftrightarrow$'' between \textit{\textbf{s}} and \textit{\textbf{t}}: \textit{\textbf{s}}$\leftrightarrow$\textit{\textbf{t}}. Reflexive arrows are omitted at each state to simplify the models. Each arrow is labelled with a numerical integer $n \in \{1,2,3,4,5,6\}$ that indicates the degree of the state located on top of the arrow. For instance, if $\underset{n}{\rightarrow}$ \textit{\textbf{s}}, then the degree of state \textit{\textbf{s}} is \textit{n}. When two states are equally plausible, for instance \textit{\textbf{s}} $\underset{n}{\leftrightarrow}$ \textit{\textbf{t}}, they receive the same plausibility degree \textit{n}.  

\medskip
 Suppose the intelligence officer has to evaluate the following message $\mathsf{m}$ from a source belonging to a belligerent country: 

\begin{center}
$\mathsf{m}$ : ``\textit{My country is building 8 nuclear submarines}$"$
\end{center}

 Suppose that for evaluating the credibility of $\mathsf{m}$, the officer detains \textit{two pieces} of relevant contextual evidence $\mathsf{e}_1$ and $\mathsf{e}_2$:
\medskip

$\mathsf{e}_1$: ``\textit{The belligerent country was delivered 8 submarine hulls}$"$
\medskip

$\mathsf{e}_2$: ``\textit{The country has decided to buy new nuclear reactors}$"$
\medskip

\medskip

Here $\mathsf{e}_1$ and $\mathsf{e}_2$ form a set of contextual evidence on which the officer can define her set of prior credibility ratings $\C^{n}\mathsf{m}$. $\mathsf{e}_1$ and $\mathsf{e}_2$. Logically, eight possible states can be distinguished based on message $\mathsf{m}$ and evidence $\mathsf{e}_1$ and $\mathsf{e}_2$:

\begin{center}\small
$\textit{\textbf{s}}^{\{\mathsf{e}_1,\mathsf{e}_2,\mathsf{m}\}},\textit{\textbf{t}}^{\{\mathsf{e}_1,\mathsf{e}_2,\neg\mathsf{m}\}},\textit{\textbf{u}}^{\{\mathsf{e}_1,\neg\mathsf{e}_2,\neg\mathsf{m}\}},\textit{\textbf{v}}^{\{\neg\mathsf{e}_1,\neg\mathsf{e}_2,\neg\mathsf{m}\}},\textit{\textbf{w}}^{\{\neg\mathsf{e}_1,\mathsf{e}_2,\mathsf{m}\}},\textit{\textbf{x}}^{\{\neg\mathsf{e}_1,\mathsf{e}_2,\neg\mathsf{m}\}},\textit{\textbf{y}}^{\{\neg\mathsf{e}_1,\neg\mathsf{e}_2,\mathsf{m}\}},\textit{\textbf{z}}^{\{\mathsf{e}_1,\neg\mathsf{e}_2,\mathsf{m}\}}$ 
\end{center}

In states where at least one piece of evidence amongst evidence $\mathsf{e}_1$ and $\mathsf{e}_2$ turns out to be true, some precondition exists for making an evaluation of message $\mathsf{m}$: states \textit{\textbf{s}}, \textit{\textbf{t}}, \textit{\textbf{u}}, \textit{\textbf{w}}, \textit{\textbf{x}} and \textit{\textbf{z}}. In states \textit{\textbf{y}} and \textit{\textbf{v}}, no precondition exists for making an evaluation of $\mathsf{m}$ since both pieces of evidence $\mathsf{e}_1$ and $\mathsf{e}_2$ turn out to be false. Suppose now that the officer gives the following \textit{prior distribution} of degrees to message $\mathsf{m}$ based on $\mathsf{e}_1$ and $\mathsf{e}_2$:
 
\begin{center}
\begin{tikzpicture}[node distance=1.05cm,world/.append style={minimum
size=0.8cm}]

\node (y) [label=90:{$\neg\mathsf{e}_1,\neg\mathsf{e}_2$},label=below:$\textbf{\textit{y}}$]  {$\mathsf{m}$};
\node (v) [label=90:{$\neg\mathsf{e}_1,\neg\mathsf{e}_2$},label=below:$\textbf{\textit{v}}$,right=of y] {$\neg\mathsf{m}$};
\node (t) [label=90:{$\mathsf{e}_1,\mathsf{e}_2$},label=below:$\textbf{\textit{t}}$,right=of v] {$\neg\mathsf{m}$};
\node (u) [label=90:{$\mathsf{e}_1,\neg\mathsf{e}_2$},label=below:$\textbf{\textit{u}}$,right=of t] {$\neg\mathsf{m}$};
\node (x) [label=90:{$\neg\mathsf{e}_1,\mathsf{e}_2$},label=below:$\textbf{\textit{x}}$,right=of u] {$\neg\mathsf{m}$};
\node (w) [label=90:{$\neg\mathsf{e}_1,\mathsf{e}_2$},label=below:$\textbf{\textit{w}}$,right=of x] {$\mathsf{m}$};
\node (z) [label=90:{$\mathsf{e}_1,\neg\mathsf{e}_2$},label=below:$\textbf{\textit{z}}$,right=of w] {$\mathsf{m}$};
\node (s) [label=90:{$\mathsf{e}_1,\mathsf{e}_2$},label=below:$\textbf{\textit{s}}$,right=of z] {$\mathsf{m}$};

\path[<->] (y) edge node[below] {6} (v);

\path[->] (v) edge node[below] {5} (t);

\path[->] (t) edge node[below] {4} (u);

\path[->] (u) edge node[below] {3} (x);

\path[<->] (x) edge node[below] {3} (w);

\path[->] (w) edge node[below] {2} (z);

\path[->] (z) edge node[below] {1} (s);

\end{tikzpicture}
\end{center}

Recall that the numbers below each arrow (for instance, $\underset{4}{\rightarrow}$ \textit{\textbf{u}}) indicate the plausibility of the state that is located on the top of the arrow (state \textit{\textbf{u}} in that case). More precisely, these numbers indicate the credibility rating of the message that is \textit{true} or \textit{false} at the state on top of the arrow. The model shows that message $\mathsf{m}$ is rated 1 at state \textit{\textbf{s}} whereas the negation of $\mathsf{m}$, namely $\neg\mathsf{m}$, is rated 5 at state \textit{\textbf{t}}. These ratings depend on the pieces of evidence $\mathsf{e}$ that turn out to be \textit{true} (or \textit{false}) when the message itself turns out to be \textit{true} (or \textit{false}).  
\medskip

As we have said, the set composed of contextual evidence, here $\mathsf{e}_1$ and $\mathsf{e}_2$, plays the role of a Kratzerian \textit{ordering source} \citep{Kratzer1981,Kratzer1991}. Possible states are ordered along the relation $\leqslant$ depending on whether or not they satisfy evidence $\mathsf{e}_1$ and/or $\mathsf{e}_2$. More precisely, those two pieces of evidence induce a plausibility ordering over the possible states since the more states satisfy pieces of evidence that are consistent with $\mathsf{m}$, the more these state are judged as plausible by the officer. On the contrary, the more the states satisfy pieces of evidence that are consistent with $\neg\mathsf{m}$, the more these states are judged as implausible by the officer.\footnote{A difference with Kratzer's framework is that in our own proposal, pieces of evidence from $\mathsf{E}$ do not have the same weight. For instance, evidence $\mathsf{e}_1$ has more weight than evidence $\neg\mathsf{e}_1$ since message $\mathsf{m}$ gets a credibility degree of 2 based on the evidence set $\{\mathsf{e}_1,\neg\mathsf{e}_2\}$ but a lower, and even uncertain, credibility degree of 3 based on the evidence set $\{\neg \mathsf{e}_1,\mathsf{e}_2\}$.} 

\medskip
When determining a credibility distribution for message $\mathsf{m}$, it seems reasonable that \textit{if} the officer gives the degree of credibility 1 to message $\mathsf{m}$ based on the evidence set $\{\mathsf{e}_1,\mathsf{e}_2\}$, the same officer will give, \textit{by symmetry}, a degree of credibility that is very low to $\neg\mathsf{m}$ based on the same evidence set, which turns out to be degree 5 in that case. A similar reasoning applies to all the degrees of credibility. The different ratings and their evidence sets are presented in Table \ref{tab:prior}.

\medskip

\begin{table}[!htbp]
  \centering
  
    \begin{tabular}{|c|c|c|}
    \hline
   
    \ \cellcolor{kugray5}\textbf{State(s)} & \cellcolor{kugray5}\textbf{Set of Evidence} & \cellcolor{kugray5}\textbf{Prior Rating} $\C^{n}\mathsf{m}$\\
   
    \hline
    
\textit{\textbf{s}} & $\{\mathsf{e}_1,\mathsf{e}_2\}$ & \small{$\C^{1}\mathsf{m}$} \\\hline

  \textit{\textbf{z}} & $\{\mathsf{e}_1,\neg\mathsf{e}_2\}$ & \small{$\C^{2}\mathsf{m}$} \\\hline

  \textit{\textbf{x}},\textit{\textbf{w}} & $\{\neg\mathsf{e}_1,\mathsf{e}_2\}$ & \small{$\neg\C^{3}\mathsf{m}\wedge
   \neg\C^{3}\neg\mathsf{m}$} \\ \hline

  \textit{\textbf{u}} & $\{\mathsf{e}_1,\neg\mathsf{e}_2\}$ & \small{$\C^{4}\neg\mathsf{m}$} \\ \hline

  \textit{\textbf{t}}  & $\{\mathsf{e}_1,\mathsf{e}_2\}$ & \small{$\C^{5}\neg\mathsf{m}$} \\\hline

  \textit{\textbf{y}},\textit{\textbf{v}}  & $\{\neg\mathsf{e}_1,\neg\mathsf{e}_2\}$ & \small{$\neg\C^{6}\mathsf{m}\wedge
   \neg\C^{6}\neg\mathsf{m}$} 
      \\\hline

    \end{tabular}%
\caption{Evidence-Based \textit{Prior} Ratings for Message $\mathsf{m}$.\label{tab:prior}}
\end{table}%


\medskip

 Now that we have matched doctrinal credibility ratings with credibility operators in \Logic, we turn to reliability ratings to provide similar clauses. But the correspondence between reliability ratings and updates of degrees is not as straightforward as before. Unlike for credibility, reliability ratings cannot be defined from literal readings of their intended descriptions.

\subsection{Rating Reliability by Updating Credibility Degrees}
\label{ssec:rel}

\subsubsection{Defining Update Rules for Credibility Degrees}
 The general semantic clause for operation $[\R^\pm\varphi]\psi$ in model $\mathds{S}$ is:

\begin{itemize}
\item[•]$\mathds{S}, \textit{\textbf{s}} \models [\R^\pm\varphi]\psi$\ \textit{iff}\ $\mathds{S}^{^{[\R^\pm\varphi]}}, \textit{\textbf{s}} \models \psi$.
\end{itemize}

\medskip
 Updated models $\mathds{S}^{^{[\R^\pm\varphi]}}$ are obtained from $\mathds{S}$ in the following way:

\smallskip
\begin{center}
$\mathds{S}^{^{[\R^\pm\varphi]}} \ =\ \ \langle \ \mathcal{S}^{^{[\R^\pm\varphi]}}, \leqslant^{^{[\R^\pm\varphi]}},\ \parallel .\parallel^{^{[\R^\pm\varphi]}} \rangle$
\end{center}

 In models $\mathds{S}^{^{[\R^\pm\varphi]}}$, the sets of states $\mathcal{S}^{^{[\R^\pm\varphi]}}$ and valuation maps $\parallel .\parallel^{^{[\R^\pm\varphi]}}$ are strictly identical to the initial set of states $\mathcal{S}$ and valuation map $\parallel .\parallel$ from $\mathds{S}$. The crucial aspects of $\mathds{S}^{^{[\R^\pm\varphi]}}$ are the plausibility preorders $\leqslant^{^{[\R^\pm\varphi]}}$. For all $\pm \in \{$A,B,C,D,E,F$\}$, each operation $\R^\pm$ with $\varphi$ induces a specific change on the ranking of states depending on whether these states satisfy formula $\varphi$ or not.     

\medskip
First of all, it is important to distinguish two \textit{kinds of situations}. There are situations in which formula $\varphi$ is evaluable at the state \textit{\textbf{s}} under consideration in the current plausibility ordering $\leqslant$: 1 $\le$ \textit{\textbf{dg}}(\textit{\textbf{s}}) < 6. But there are also situations in which formula $\varphi$ is not evaluable at the state \textit{\textbf{s}} under consideration in the current plausibility ordering $\leqslant$: \textit{\textbf{dg}}(\textit{\textbf{s}}) = 6. Making such a distinction is crucial since in the first case (i.e. when 1 $\le$ \textit{\textbf{dg}}(\textit{\textbf{s}}) < 6), the credibility of $\varphi$ in the new ordering $\leqslant^{^{[\R^\pm\varphi]}}$ will depend \textit{both} on the prior credibility of the formula at state \textit{\textbf{s}} \textit{and} on the level of reliability $\pm$ of its source at the same state. In the second case (viz. \textit{\textbf{dg}}(\textit{\textbf{s}}) = 6), the credibility of formula $\varphi$ in the updated ordering $\leqslant^{^{[\R^\pm\varphi]}}$ will depend \textit{only} on the reliability of its source at state \textit{\textbf{s}}. 

\medskip
 In both situations, however, we write $\textit{\textbf{dg}}(\textit{\textbf{s}})^{\R^\pm\varphi}$ for the \textit{posterior degree} of state \textit{\textbf{s}} obtained by exerting operation $\pm$ with formula $\varphi$ on the \textit{prior degree} $\textit{\textbf{dg}}(\textit{\textbf{s}})$ of state \textit{\textbf{s}}. To ensure that new degrees $\textit{\textbf{dg}}^{\R^\pm\varphi}$ remain  within the bounds of set $\{1,2,3,4,5\}$, we introduce the function \textit{\texttt{Cut}}($x$) that will be applied in the definition of updates.\footnote{This technical device is inspired by \cite{Aucher2004}.}  

\smallskip

\begin{table}[!htbp]
  \centering
  
\begin{tabular}{ccc}

$\textit{\texttt{Cut}}(x)$ & = & 

$
\begin{cases}
x & \text{\textit{if} $1 \leqslant x \leqslant 5$} \\
1 & \text{\textit{if} $x < 1$} \\
5 & \text{\textit{if} $x > 5$}
\end{cases}
$
\end{tabular}
\end{table}

 Let us start out by defining the operations $\R^\pm$ when formula $\varphi$ is evaluable at the state \textit{\textbf{s}} under consideration: \textit{\textbf{dg}}(\textit{\textbf{s}}) < 6. Operation $\R^\text{A}\varphi$ is defined as the update in which all $\varphi$-states of the initial plausibility ordering $\leqslant$ gain 3 ranks in the new ordering $\leqslant^{^{[\R^\text{A}\varphi]}}$ while all $\neg\varphi$-states of the initial plausibility ordering $\leqslant$ lose 3 ranks in the new ordering $\leqslant^{^{[\R^\text{A}\varphi]}}$. In terms of credibility degrees, this means that the degrees of $\varphi$-states \textit{decrease by} 3 while degrees of $\neg\varphi$-states \textit{increase by} 3. Accordingly, operation $\R^\text{A}\varphi$ is defined by:

\begin{table}[!htbp]
  \centering
  
\begin{tabular}{ccc}
 $\textit{\textbf{dg}}(\textit{\textbf{s}})^{\R^\text{A}\varphi}$ & = & 

$
\begin{cases}
\ \textit{\texttt{Cut}}(\textit{\textbf{dg}}(\textit{\textbf{s}}) - 3) & \text{\textit{if}}\ \mathds{S}, \textit{\textbf{s}} \models \varphi. \\
                                   
\ \textit{\texttt{Cut}}(\textit{\textbf{dg}}(\textit{\textbf{s}}) + 3) & \text{\textit{if}}\ \mathds{S}, \textit{\textbf{s}} \models \neg\varphi. \\                 
\end{cases}$
\end{tabular}
\end{table}

 Operation $\R^\text{B}\varphi$ is the update according to which all $\varphi$-states of the initial plausibility ordering gain 2 ranks in the new ordering $\leqslant^{^{[\R^\text{B}\varphi]}}$ (their degrees \textit{decrease by} 2) while all $\neg\varphi$-states of the initial plausibility ordering lose 2 ranks in the new ordering $\leqslant^{^{[\R^\text{B}\varphi]}}$. Numerically, $\R^\text{B}\varphi$ is defined by:

\begin{table}[!htbp]
  \centering
  
\begin{tabular}{ccc}

$\textit{\textbf{dg}}(\textit{\textbf{s}})^{\R^\text{B}\varphi}$ & = & 

$
\begin{cases}
                                   \ \textit{\texttt{Cut}}(\textit{\textbf{dg}}(\textit{\textbf{s}}) - 2) & \text{\textit{if}}\ \mathds{S}, \textit{\textbf{s}} \models \varphi. \\
                                   
                                   \ \textit{\texttt{Cut}}(\textit{\textbf{dg}}(\textit{\textbf{s}}) + 2) & \text{\textit{if}}\ \mathds{S}, \textit{\textbf{s}} \models \neg\varphi. \\
                          
\end{cases}$
\end{tabular}
\end{table}

 Operation $\R^\text{C}\varphi$ is the update according to which all $\varphi$-states of the initial ordering gain one rank in the new ordering $\leqslant^{^{[\R^\text{C}\varphi]}}$ while all $\neg\varphi$-states of the initial ordering lose 1 rank in the new ordering $\leqslant^{^{[\R^\text{C}\varphi]}}$. Numerically, operation $\R^\text{C}\varphi$ is given by:

\begin{table}[!htbp]
  \centering
  
\begin{tabular}{ccc}

$\textit{\textbf{dg}}(\textit{\textbf{s}})^{\R^\text{C}\varphi}$ & = & 

$
  \begin{cases}
                                   \ \textit{\texttt{Cut}}(\textit{\textbf{dg}}(\textit{\textbf{s}}) - 1) & \text{\textit{if}}\ \mathds{S}, \textit{\textbf{s}} \models \varphi. \\
                                   
                                   \ \textit{\texttt{Cut}}(\textit{\textbf{dg}}(\textit{\textbf{s}}) + 1) & \text{\textit{if}}\ \mathds{S}, \textit{\textbf{s}} \models \neg\varphi. \\
                          
  \end{cases}$
  
\end{tabular}
\end{table}

 Contrary to the operations $\R^\text{A}\varphi$ to $\R^\text{C}\varphi$ that have a promoting effect on $\varphi$-states, operation $\R^\text{D}\varphi$ is the update according to which all $\varphi$-states of the initial ordering lose 1 rank in the new ordering $\leqslant^{^{[\R^\text{D}\varphi]}}$ (their degrees \textit{increase by} 1) while all $\neg\varphi$-states of the initial ordering gain 1 rank in the new ordering $\leqslant^{^{[\R^\text{D}\varphi]}}$ (their degrees \textit{decrease by} 1). In terms of plausibility degrees, operation $\R^\text{D}\varphi$ is defined by:

\begin{table}[!htbp]
  \centering
  
\begin{tabular}{ccc}

$\textit{\textbf{dg}}(\textit{\textbf{s}})^{\R^\text{D}\varphi}$ & = & 

$
  \begin{cases}
                                   \ \textit{\texttt{Cut}}(\textit{\textbf{dg}}(\textit{\textbf{s}}) + 1) & \text{\textit{if}}\ \mathds{S}, \textit{\textbf{s}} \models \varphi. \\
                                   
                                   \ \textit{\texttt{Cut}}(\textit{\textbf{dg}}(\textit{\textbf{s}}) - 1) & \text{\textit{if}}\ \mathds{S}, \textit{\textbf{s}} \models \neg\varphi. \\
                          
  \end{cases}$
\end{tabular}
\end{table}

 Operation $\R^\text{E}\varphi$ is the update such that all $\varphi$-states of the initial ordering lose 2 ranks in the new ordering $\leqslant^{^{[\R^\text{E}\varphi]}}$ while all $\neg\varphi$-states of the initial ordering gain 2 ranks in the new ordering $\leqslant^{^{[\R^\text{E}\varphi]}}$. Numerically, then, operation $\R^\text{E}\varphi$ is defined by:  

\begin{table}[!htbp]
  \centering
  
\begin{tabular}{ccc}

$\textit{\textbf{dg}}(\textit{\textbf{s}})^{\R^\text{E}\varphi}$ & = &  
$
  \begin{cases}
                                   \ \textit{\texttt{Cut}}(\textit{\textbf{dg}}(\textit{\textbf{s}}) + 2) & \text{\textit{if}}\ \mathds{S}, \textit{\textbf{s}} \models \varphi. \\
                                   
                                   \ \textit{\texttt{Cut}}(\textit{\textbf{dg}}(\textit{\textbf{s}}) - 2) & \text{\textit{if}}\ \mathds{S}, \textit{\textbf{s}} \models \neg\varphi. \\
                          
\end{cases}$
\end{tabular}
\end{table}

 Contrary to all the operations above, operation $\R^\text{F}\varphi$ leaves the initial plausibility ordering $\leqslant$ as it is, — no matter if formula $\varphi$ is true or false at the state where the operation is performed. Accordingly, operation $\R^\text{F}\varphi$ is defined by:

\begin{table}[!htbp]
  \centering
  
\begin{tabular}{cccc}
&$\textit{\textbf{dg}}(\textit{\textbf{s}})^{\R^\text{F}\varphi}$ & = & $\textit{\textbf{dg}}(\textit{\textbf{s}})\ \ \ \textit{if} \ \mathds{S}, \textit{\textbf{s}} \models \varphi$\ \textit{or if} \ $\mathds{S}, \textit{\textbf{s}} \models \neg\varphi$. 
\end{tabular}
\end{table}

 So far, we have defined the $\R^\pm$-operations when the degree of the evaluation state \textit{\textbf{s}} is less than 6. We have focused our attention on cases in which evidence already exists at state \textit{\textbf{s}} for making an evaluation of formula $\varphi$. But what if all the evidence turns out to be false at the state \textit{\textbf{s}} such that the credibility of formula $\varphi$ cannot be judged? In that case, the officer's evaluation will be based \textit{only} on the reliability of the source. In other words, posterior credibility degrees $\textit{\textbf{dg}}(\textit{\textbf{s}})^{\R^\pm\varphi}$ will stricly reduce to the (level of) reliability $\R^\pm$ of the source at state \textit{\textbf{s}} \citep[see][200]{Samet1975}. For this reason, we associate a \textit{fixed degree} to all the $\R^\pm$-operations performed at states of degree 6. These degrees will reflect the \textit{gain} or \textit{loss} of credibility resulting from the intended operation.  

\medskip
 In case formula $\varphi$ is not evaluable at the state \textit{\textbf{s}} under consideration (\textit{\textbf{dg}}(\textit{\textbf{s}}) = 6), operation $\R^\text{A}\varphi$ is the update according to which all $\varphi$-states of the initial ordering $\leqslant$ go to the \textit{first} rank in the new ordering $\leqslant^{^{[\R^\text{A}\varphi]}}$ while all $\neg\varphi$-states of the initial ordering $\leqslant$ go to the \textit{fifth} rank in the new ordering $\leqslant^{^{[\R^\text{A}\varphi]}}$. Numerically, operation $\R^\text{A}\varphi$ is defined by:

\begin{table}[!htbp]
  \centering
  
\begin{tabular}{ccc}

$\textit{\textbf{dg}}(\textit{\textbf{s}})^{\R^\text{A}\varphi}$ & = &
$
  \begin{cases}
                                   \ 1 & \text{\textit{if}}\ \mathds{S}, \textit{\textbf{s}} \models \varphi. \\
                                   
                                   \ 5 & \text{\textit{if}}\ \mathds{S}, \textit{\textbf{s}} \models \neg\varphi. \\
                          
\end{cases}$
\end{tabular}
\end{table}

 When formula $\varphi$ is not assessable at state \textit{\textbf{s}}, operation $\R^\text{B}\varphi$ makes the states of the initial ordering $\leqslant$ go to the \textit{second} rank in the new ordering $\leqslant^{^{[\R^\text{B}\varphi]}}$ if these states satisfy formula $\varphi$, and to the \textit{fourth} rank in the new ordering $\leqslant^{^{[\R^\text{B}\varphi]}}$ if these states satisfy $\neg\varphi$. In terms of credibility degrees, operation $\R^\text{B}\varphi$ is defined by:

\begin{table}[!htbp]
  \centering
  
\begin{tabular}{ccc}

$\textit{\textbf{dg}}(\textit{\textbf{s}})^{\R^\text{B}\varphi}$ & = &
$
\begin{cases}
\ 2, & \text{\textit{if}}\ \mathds{S}, \textit{\textbf{s}} \models \varphi. \\

\ 4, & \text{\textit{if}}\ \mathds{S}, \textit{\textbf{s}} \models \neg\varphi. 
\end{cases}$
\end{tabular}
\end{table}

 Operation $\R^\text{C}\varphi$ is the update according to which all the states of the initial ordering $\leqslant$ go to the \textit{third} rank in the new ordering $\leqslant^{^{[\R
^\text{C}\varphi]}}$ no matter whether these states satisfy formula $\varphi$ or satisfy formula $\neg\varphi$. Numerically, operation $\R^\text{C}\varphi$ is defined by:

\begin{table}[!htbp]
  \centering
  
\begin{tabular}{ccccccc}

& & & & $\textit{\textbf{dg}}(\textit{\textbf{s}})^{\R^\text{C}\varphi}$ & = & 3 $\textit{if} \ \mathds{S}, \textit{\textbf{s}} \models \varphi\ \textit{or if} \ \mathds{S}, \textit{\textbf{s}} \models \neg\varphi$. 
\end{tabular}
\end{table}

   Operation $\R^\text{D}\varphi$ makes the states of the initial ordering $\leqslant$ go to the \textit{fourth} rank in the new ordering $\leqslant^{^{[\R^\text{D}\varphi]}}$ if these states satisfy formula $\varphi$, and to the \textit{second} rank in the new ordering $\leqslant^{^{[\R^\text{D}\varphi]}}$ if these states satisfy $\neg\varphi$. In terms of credibility degrees, operation $\R^\text{D}\varphi$ is defined by:

\begin{table}[!htbp]
  \centering
  
\begin{tabular}{ccc}

$\textit{\textbf{dg}}(\textit{\textbf{s}})^{\R^\text{D}\varphi}$ & = & 
  $\begin{cases}
                                   \ 4 & \text{\textit{if}}\ \mathds{S}, \textit{\textbf{s}} \models \varphi. \\
                                   
                                   \ 2 & \text{\textit{if}}\ \mathds{S}, \textit{\textbf{s}} \models \neg\varphi. \\
                          
  \end{cases}$
\end{tabular}
\end{table}

 Operation $\R^\text{E}\varphi$ is the update according to which all $\varphi$-states of the initial ordering $\leqslant$ go to the \textit{fifth} rank in the new ordering $\leqslant^{^{[\R^\text{E}\varphi]}}$ while all $\neg\varphi$-states of the initial ordering $\leqslant$ go to the \textit{first} rank in the new ordering $\leqslant^{^{[\R^\text{E}\varphi]}}$. In terms of degrees, operation $\R^\text{E}\varphi$ is defined by:

\begin{table}[!htbp]
  \centering
  
\begin{tabular}{ccc}

$\textit{\textbf{dg}}(\textit{\textbf{s}})^{\R^\text{E}\varphi}$ & = &
$\begin{cases}
                                   \ 5 & \text{\textit{if}}\ \mathds{S}, \textit{\textbf{s}} \models \varphi. \\
                                   
                                   \ 1 & \text{\textit{if}}\ \mathds{S}, \textit{\textbf{s}} \models \neg\varphi. \\
                          
  \end{cases}$

\end{tabular}
\end{table}

 Finally, operation $\R^\text{F}\varphi$ is the update according to which all the states of the initial ordering $\leqslant$ go to the \textit{sixth} rank in the new ordering $\leqslant^{^{[\R
^\text{F}\varphi]}}$, — no matter if these states satisfy formula $\varphi$ or satisfy formula $\neg\varphi$. Numerically, operation $\R^\text{F}\varphi$ is defined by:

\begin{table}[!htbp]
  \centering
  
\begin{tabular}{cccccccc}
&&&&&
$\textit{\textbf{dg}}(\textit{\textbf{s}})^{\R^\text{F}\varphi}$ & = & 6\ \ \ \textit{if} \ $\mathds{S}, \textit{\textbf{s}} \models \varphi$\ \textit{or if} \ $\mathds{S}, \textit{\textbf{s}} \models \neg\varphi$. 

\end{tabular}
\end{table}

The scoring operations we proposed for updating credibility degrees may seem arbitrary to some extent. We may wonder why their definitions should be based on adding or substracting degrees, and why we should add or remove 1 to 3 degrees in such and such case instead of other quantities. First of all, our proposal is consistent with empirical observations: in \Logic, credibility is the underlying semantic dimension of evaluation while reliability only plays an adjustement role. In addition to that, the update rules defined for operators $\R^\pm$ are in line with other observations made by \cite{mandel2022meta}, in particular on the positive versus negative groupings of reliability operators. Consistent with the positive directionality of ratings \textbf{A} to \textbf{C}, operators $\R^\text{A}$ to $\R^\text{C}$ lead to substract degrees from their targeted state if they satisfy the intended formula $\varphi$. By contrast, operators $\R^\text{D}$ and $\R^\text{E}$ add degrees to their targeted states now in line with the negative directionality of reliablility ratings \textbf{D} and \textbf{E}. Operator $\R^\text{F}$ leaves the initial degree of the states as they were initially as it happens to be the case when the reliability of the source cannot be judged (rating \textbf{F}). As a matter of fact, we will see later that our setting also turns out to be consistent with the proposals made to enforce the \textit{facts versus interpretations recommendation}. 

\subsubsection{Matching Update Rules with Reliability Ratings}
\label{sssec:match-rel}

Before doing so, let us match the various reliability operators with the doctrinal reliability ratings ranging from \textbf{A} to \textbf{F}. Intuitively, the more reliable officers judge a source of information to be, the more they will favor contextual evidence that is consistent with the message from this source. On the contrary, the less reliable officers judge a source to be, the more they will favor contextual evidence inconsistent with the message from this source. When officers are unable to assess the reliability of a source, they keep credibility ratings as they are. Our proposal to match reliability operators $\R^\pm$ with doctrinal reliability ratings is based on those intuitions. In model $\mathds{S}$, \textit{favoring} or \textit{dismissing} pieces of evidence that are consistent or inconsistent with a message amounts to \textit{decreasing} or \textit{increasing} the rank of their corresponding state.    

\medskip
 The criteria for deciding whether sources are \textit{reliable}, \textit{unreliable} or \textit{unassessable}, depends on their \textit{informational pedigree}. This means that levels of reliability that officers may put into sources are determined by the \textit{truth} or \textit{falsity} of the information they have provided in the past. That being determined, officers know how much they should promote or dismiss pieces of evidence that support or contradict the message $\mathsf{m}$ they are evaluating. 

\medskip
 Following doctrinal descriptions (see Table \ref{tab:rel}), sources are classified as ``Completely Reliable$"$ and rated \textbf{A} if they \textit{have a history of complete reliability} such that there is \textit{no doubt about their authenticity and trustworthiness}. In such conditions, the officer is \textit{strongly justified} to promote the states \textit{\textbf{s}} satisfying some evidence $\mathsf{e}$ that is consistent with message $\mathsf{m}$ being true and to dismiss the states \textit{\textbf{s}} satisfying some evidence $\mathsf{e}$ that are consistent with $\mathsf{m}$ being false. Accordingly, we suggest matching rating \textbf{A} with reliability operator $\R^\text{A}$ from \Logic. In case message $\mathsf{m}$ was already evaluable at state \textit{\textbf{s}} (because \textit{some distinct evidence} $\mathsf{e}$ exists for assessing $\mathsf{m}$, i.e. 1 $\le$ \textit{\textbf{dg}}(\textit{\textbf{s}}) < 6), the degree of state \textit{\textbf{s}} \textit{decreases by} 3 if \textit{\textbf{s}} is consistent with $\mathsf{m}$ being true: $\textit{\textbf{dg}}(\textit{\textbf{s}})^{\R^\text{A}\varphi}\ \ =\ \ \textit{\texttt{Cut}}(\textit{\textbf{dg}}(\textit{\textbf{s}}) - 3)$ \textit{if} $\mathds{S}, \textit{\textbf{s}} \models \mathsf{m}$. On the contrary, the degree of state \textit{\textbf{s}} \textit{increases by} 3 if \textit{\textbf{s}} is consistent with $\mathsf{m}$ being false: $\textit{\textbf{dg}}(\textit{\textbf{s}})^{\R^\text{A}\varphi} = \textit{\texttt{Cut}}(\textit{\textbf{dg}}(\textit{\textbf{s}}) + 3)$ \textit{if} $\mathds{S}, \textit{\textbf{s}} \models \neg\mathsf{m}$. In case message $\mathsf{m}$ was not already evaluable at state \textit{\textbf{s}} (because \textit{no distinct piece of evidence} $\mathsf{e}$ exists for assessing $\mathsf{m}$, i.e. \textit{\textbf{dg}}(\textit{\textbf{s}}) = 6), the degree of state \textit{\textbf{s}} becomes 1 if \textit{\textbf{s}} satisfies $\mathsf{m}$: $\textit{\textbf{dg}}(\textit{\textbf{s}})^{\R^\text{A}\varphi} = 1$ \textit{if} $\mathds{S}, \textit{\textbf{s}} \models \mathsf{m}$, and becomes 5 if \textit{\textbf{s}} satisfies $\neg\mathsf{m}$: $\textit{\textbf{dg}}(\textit{\textbf{s}})^{\R^\text{A}\varphi} = 5$ \textit{if} $\mathds{S}, \textit{\textbf{s}} \models \neg\mathsf{m}$. 

\medskip
 This clause expresses the strongest trusting attitude the officer may have towards the intelligence source. But this strength can be relaxed to express weaker trusting attitudes. We know that officers classify sources as ``Usually Reliable$"$ and rate them \textbf{B} if they \textit{have a history of valid information most of the time} such that there is \textit{minor doubt about their authenticity and trustworthiness}. Now officers are \textit{moderately justified} to promote $\mathsf{m}$-consistent states on top of their plausibility ordering. This level can be captured by operator $\R^\text{B}$ from \Logic. In case message $\mathsf{m}$ was already evaluable at state \textit{\textbf{s}} (i.e. 1 $\le$ \textit{\textbf{dg}}(\textit{\textbf{s}}) < 6), the degree of state \textit{\textbf{s}} \textit{decreases by} 2 if \textit{\textbf{s}} is consistent with $\mathsf{m}$ being true: $\textit{\textbf{dg}}(\textit{\textbf{s}})^{\R^\text{B}\varphi} = \textit{\texttt{Cut}}(\textit{\textbf{dg}}(\textit{\textbf{s}}) - 2)$ \textit{if} $\mathds{S}, \textit{\textbf{s}} \models \mathsf{m}$, and \textit{increases by} 2 if \textit{\textbf{s}} is consistent with $\mathsf{m}$ being false: $\textit{\textbf{dg}}(\textit{\textbf{s}})^{\R^\text{B}\varphi} = \textit{\texttt{Cut}}(\textit{\textbf{dg}}(\textit{\textbf{s}}) + 2)$ \textit{if} $\mathds{S}, \textit{\textbf{s}} \models \neg\mathsf{m}$. But in case message $\mathsf{m}$ was not already evaluable at state \textit{\textbf{s}} (i.e. \textit{\textbf{dg}}(\textit{\textbf{s}}) = 6), the degree of state \textit{\textbf{s}} becomes 2 if it satisfies $\mathsf{m}$: $\textit{\textbf{dg}}(\textit{\textbf{s}})^{\R^\text{B}\varphi} = 2\ $\textit{if} $\mathds{S}, \textit{\textbf{s}} \models \mathsf{m}$, and becomes 4 if it satisfies $\neg\mathsf{m}$: $\textit{\textbf{dg}}(\textit{\textbf{s}})^{\R^\text{B}\varphi} = 4$ \textit{if} $\mathds{S}, \textit{\textbf{s}} \models \neg\mathsf{m}$.

\medskip
 By contrast, the description provided with rating ``Fairly Reliable$"$ is more mixed: sources are rated \textbf{C} if they \textit{have provided valid information in the past} but there is some \textit{doubt concerning their authenticity and trustworthiness}. Because of this, officers should make a \textit{minor revision} of the plausibility ordering. This corresponds to reliability operator $\R^\text{C}$ in \Logic. In case message $\mathsf{m}$ was already evaluable at the state \textit{\textbf{s}} under consideration (i.e. 1 $\le$ \textit{\textbf{dg}}(\textit{\textbf{s}}) < 6), the degree of state \textit{\textbf{s}} \textit{decreases by} 1 if \textit{\textbf{s}} is consistent with $\mathsf{m}$ being true: $\textit{\textbf{dg}}(\textit{\textbf{s}})^{\R^\text{C}\varphi} = \textit{\texttt{Cut}}(\textit{\textbf{dg}}(\textit{\textbf{s}}) - 1)$ \textit{if} $\mathds{S}, \textit{\textbf{s}} \models \mathsf{m}$, and \textit{increases by} 1 if \textit{\textbf{s}} is consistent with $\mathsf{m}$ being false: $\textit{\textbf{dg}}(\textit{\textbf{s}})^{\R^\text{C}\varphi} =  \textit{\texttt{Cut}}(\textit{\textbf{dg}}(\textit{\textbf{s}}) + 1)$ \textit{if} $\mathds{S}, \textit{\textbf{s}} \models \neg\mathsf{m}$. In case message $\mathsf{m}$ was not already evaluable at state \textit{\textbf{s}} (i.e. \textit{\textbf{dg}}(\textit{\textbf{s}}) = 6), the degree of state \textit{\textbf{s}} becomes 3 no matter whether it satisfies $\mathsf{m}$ or satisfies $\neg\mathsf{m}$: $\textit{\textbf{dg}}(\textit{\textbf{s}})^{\R^\text{C}\varphi} = 3$ \textit{if} $\mathds{S}, \textit{\textbf{s}} \models \mathsf{m}$ \textit{or if} $\mathds{S}, \textit{\textbf{s}} \models \neg\mathsf{m}$.

\medskip
 According to the doctrine, sources are classified as ``Not Usually Reliable$"$ and rated \textbf{D} if there is \textit{significant doubt concerning their authenticity and trustworthiness} but they \textit{have provided valid information in the past}. Expression ``\textit{significant doubt}$"$ clearly indicates that states which satisfy inconsistent evidence with $\mathsf{m}$ become more plausible than states satisfying evidence consistent with $\mathsf{m}$. But this promotion is minimized by the fact that the sources delivered valid information in the past. So the officer is \textit{barely justified} in promoting $\mathsf{m}$-inconsistent states on top of his or her plausibility ordering. The corresponding operation is $\R^\text{D}$ in language \Logic. In case message $\mathsf{m}$ was already evaluable at state \textit{\textbf{s}}, this operator makes the degree of \textit{\textbf{s}} \textit{decreases by} 1 if \textit{\textbf{s}} is consistent with $\mathsf{m}$ being false: $\textit{\textbf{dg}}(\textit{\textbf{s}})^{\R^\text{D}\varphi} =  \textit{\texttt{Cut}}(\textit{\textbf{dg}}(\textit{\textbf{s}}) - 1)$ \textit{if} $\mathds{S}, \textit{\textbf{s}} \models \neg\mathsf{m}$, and \textit{increases by} 1 if \textit{\textbf{s}} is consistent with $\mathsf{m}$ being true: $\textit{\textbf{dg}}(\textit{\textbf{s}})^{\R^\text{D}\varphi} =  \textit{\texttt{Cut}}(\textit{\textbf{dg}}(\textit{\textbf{s}}) + 1)$ \textit{if} $\mathds{S}, \textit{\textbf{s}} \models \mathsf{m}$. In case message $\mathsf{m}$ was not already evaluable at state \textit{\textbf{s}}, the degree of state \textit{\textbf{s}} becomes 1 if it satisfies $\neg\mathsf{m}$: $\textit{\textbf{dg}}(\textit{\textbf{s}})^{\R^\text{D}\varphi} = 1$ \textit{if} $\mathds{S}, \textit{\textbf{s}} \models \neg\mathsf{m}$, and becomes 5 if it satisfies $\mathsf{m}$: $\textit{\textbf{dg}}(\textit{\textbf{s}})^{\R^\text{D}\varphi} = 5$ \textit{if} $\mathds{S}, \textit{\textbf{s}} \models \mathsf{m}$.

\medskip
 Concerning the rating ``Unreliable$"$, the doctrine states that sources are classified \textbf{E} in case they are \textit{lacking in authenticity and trustworthiness} and have an \textit{history of invalid information}. Now officers are \textit{highly justified} to promote $\mathsf{m}$-inconsistent states on top of their plausibility ordering. But they are not as justified as they were to promote $\mathsf{m}$-consistent states in case of rating \textbf{A}. In the latter case, the doctrinal description indicates that sources have an history of \textit{complete} reliability. In case of rating \textbf{E}, sources are not trustworthy and have provided invalid information in the past but they are not described as \textit{completely} unreliable. For this reason, officers are not strongly justified to promote $\mathsf{m}$-inconsistent states on top of the plausibility ordering. This moderate attitude can be captured by operator $\R^\text{E}$ in \Logic. In case message $\mathsf{m}$ was already evaluable at state \textit{\textbf{s}}, this operator makes the degree of \textit{\textbf{s}} \textit{decreases by} 2 if \textit{\textbf{s}} is consistent with $\mathsf{m}$ being false: $\textit{\textbf{dg}}(\textit{\textbf{s}})^{\R^\text{E}\varphi} = \textit{\texttt{Cut}}(\textit{\textbf{dg}}(\textit{\textbf{s}}) - 2)$ \textit{if} $\mathds{S}, \textit{\textbf{s}} \models \neg\mathsf{m}$, and \textit{increases by} 2 if \textit{\textbf{s}} is consistent with $\mathsf{m}$ being true: $\textit{\textbf{dg}}(\textit{\textbf{s}})^{\R^\text{E}\varphi} = \textit{\texttt{Cut}}(\textit{\textbf{dg}}(\textit{\textbf{s}}) + 2)$ \textit{if} $\mathds{S}, \textit{\textbf{s}} \models \mathsf{m}$. In case message $\mathsf{m}$ was not already evaluable at state \textit{\textbf{s}}, the degree of state \textit{\textbf{s}} becomes 2 if it satisfies $\neg\mathsf{m}$: $\textit{\textbf{dg}}(\textit{\textbf{s}})^{\R^\text{E}\varphi} = 2\ $\textit{if} $\mathds{S}, \textit{\textbf{s}} \models \neg\mathsf{m}$, and becomes 4 if it satisfies $\mathsf{m}$: $\textit{\textbf{dg}}(\textit{\textbf{s}})^{\R^\text{E}\varphi} = 4$ \textit{if} $\mathds{S}, \textit{\textbf{s}} \models \mathsf{m}$. 

\medskip
 Eventually, rating \textbf{F} is ascribed when the reliability of the source ``Cannot Be Judged$"$. In this case, \textit{no basis exists for evaluating the reliability of the source}. This means that the pieces of evidence the officer has are \textit{all false} and for this reason, they do not provide a reasonable ground for adjudicating on the truth value of message $\mathsf{m}$. A corresponding operator in \Logic is $\R^\text{F}$: no matter whether the message was already evaluable at state \textit{\textbf{s}} or not, operator $\R^\text{F}$ keeps the degree of \textit{\textbf{s}} as it was if \textit{\textbf{s}} satisfies $\mathsf{m}$ or satisfies $\neg\mathsf{m}$: $\textit{\textbf{dg}}(\textit{\textbf{s}})^{\R^\text{F}\varphi} =  \textit{\textbf{dg}}(\textit{\textbf{s}})$ \textit{if} $\mathds{S}, \textit{\textbf{s}} \models \mathsf{m}$ \textit{or if} $\mathds{S}, \textit{\textbf{s}} \models \neg\mathsf{m}$. If $\mathsf{m}$ was not evaluable at state \textit{\textbf{s}} (i.e. \textit{\textbf{dg}}(\textit{\textbf{s}}) = 6), the degree of \textit{\textbf{s}} remains equal to 6 after operation $\R^\text{F}$ is performed in that case. Table \ref{tab:sumrel} sums up all the semantic clauses we have given for expressing reliability ratings from \textbf{A} to \textbf{F}.

\begin{table}[htbp]
  \centering
  
    \begin{tabular}{|c|c||c|c|}
    \hline
   \rowcolor{kugray5} & & & \\
    \rowcolor{kugray5}\textbf{Rating} & \textbf{Label} & \textbf{If} 1 $\le$ \textit{\textbf{dg}}(\textit{\textbf{s}}) < 6 &  \textbf{If} \textit{\textbf{dg}}(\textit{\textbf{s}}) = 6 \\
   \rowcolor{kugray5} & & & \\
    \hline
    & & & \\
     $\R^\text{A}\varphi$ & \small{Completely Reliable} & \mbox{\normalsize $\textit{\texttt{Cut}}(\textit{\textbf{dg}}(\textit{\textbf{s}}) - 3)$ \textit{if} $\mathds{S}, \textit{\textbf{s}} \models \varphi$} & 1 \textit{if} $\mathds{S}, \textit{\textbf{s}} \models \varphi$ \\
     
    & & \mbox{\normalsize $\textit{\texttt{Cut}}(\textit{\textbf{dg}}(\textit{\textbf{s}}) + 3)$ \textit{if} $\mathds{S}, \textit{\textbf{s}} \models \neg\varphi$} & 5 \textit{if} $\mathds{S}, \textit{\textbf{s}} \models \neg\varphi$ \\
       & & & \\\hline
    & & & \\
     $\R^\text{B}\varphi$ & \small{Usually Reliable} & \mbox{\normalsize $\textit{\texttt{Cut}}(\textit{\textbf{dg}}(\textit{\textbf{s}}) - 2)$ \textit{if} $\mathds{S}, \textit{\textbf{s}} \models \varphi$} & 2 \textit{if} $\mathds{S}, \textit{\textbf{s}} \models \varphi$ \\
     
    & &  \mbox{\normalsize $\textit{\texttt{Cut}}(\textit{\textbf{dg}}(\textit{\textbf{s}}) + 2)$ \textit{if} $\mathds{S}, \textit{\textbf{s}} \models \neg\varphi$} & 4 \textit{if} $\mathds{S}, \textit{\textbf{s}} \models \neg\varphi$ \\
     & & & \\\hline
    & & & \\
      & & \mbox{\normalsize $\textit{\texttt{Cut}}(\textit{\textbf{dg}}(\textit{\textbf{s}}) - 1)$ \textit{if} $\mathds{S}, \textit{\textbf{s}} \models \varphi$} &  \\ 
      
        $\R^\text{C}\varphi$ & \small{Fairly Reliable} & & 3 \\ 
     & & \mbox{\normalsize $\textit{\texttt{Cut}}(\textit{\textbf{dg}}(\textit{\textbf{s}}) + 1)$ \textit{if} $\mathds{S}, \textit{\textbf{s}} \models \neg\varphi$} & \\
     
    & & & \\\hline
    & & & \\
      $\R^\text{D}\varphi$ & \small{Not Usually Reliable} & \mbox{\normalsize $\textit{\texttt{Cut}}(\textit{\textbf{dg}}(\textit{\textbf{s}}) + 1)$ \textit{if} $\mathds{S}, \textit{\textbf{s}} \models \varphi$} & 4 \textit{if} $\mathds{S}, \textit{\textbf{s}} \models \varphi$ \\ 

     & & \mbox{\normalsize $\textit{\texttt{Cut}}(\textit{\textbf{dg}}(\textit{\textbf{s}}) - 1)$ \textit{if} $\mathds{S}, \textit{\textbf{s}} \models \neg\varphi$} & 2 \textit{if} $\mathds{S}, \textit{\textbf{s}} \models \neg\varphi$ \\
     & & & \\\hline
    & & & \\
      $\R^\text{E}\varphi$ & \small{Unreliable} & \mbox{\normalsize $\textit{\texttt{Cut}}(\textit{\textbf{dg}}(\textit{\textbf{s}}) + 2)$ \textit{if} $\mathds{S}, \textit{\textbf{s}} \models \varphi$} & 5 \textit{if} $\mathds{S}, \textit{\textbf{s}} \models \varphi$ \\ 
     
    & &
     \mbox{\normalsize $\textit{\texttt{Cut}}(\textit{\textbf{dg}}(\textit{\textbf{s}}) - 2)$ \textit{if} $\mathds{S}, \textit{\textbf{s}} \models \neg\varphi$} & 1 \textit{if} $\mathds{S}, \textit{\textbf{s}} \models \neg\varphi$ \\
 
         & & & \\\hline
     & & & \\
     
      $\R^\text{F}\varphi$ & \small{Cannot Be Judged} & 
\normalsize \textit{\textbf{dg}}(\textit{\textbf{s}})  & \normalsize \textit{\textbf{dg}}(\textit{\textbf{s}}) \\

    & & & \\\hline

    \end{tabular}%
\caption{Semantic Conditions for Reliability Updates.}
\label{tab:sumrel}
\end{table}%

\subsubsection{A follow-up on the submarines example}
\label{sssec:case2}
 Let us go back to the submarine example we used as a modelling case study in subsection \ref{sssec:case1}. 
 Recall that the model we proposed for a possible prior distribution of ratings for $\mathsf{m}$ based on $\mathsf{e}_1$ and $\mathsf{e}_2$ was: 

\begin{center}
\begin{tikzpicture}[node distance=1.05cm,world/.append style={minimum
size=0.8cm}]
\node (y) [label=90:{$\neg\mathsf{e}_1,\neg\mathsf{e}_2$},label=below:$\textbf{\textit{y}}$,left=of v]  {$\mathsf{m}$};
\node (v) [label=90:{$\neg\mathsf{e}_1,\neg\mathsf{e}_2$},label=below:$\textbf{\textit{v}}$,right=of y] {$\neg\mathsf{m}$};
\node (t) [label=90:{$\mathsf{e}_1,\mathsf{e}_2$},label=below:$\textbf{\textit{t}}$,right=of v] {$\neg\mathsf{m}$};
\node (u) [label=90:{$e,\neg\mathsf{e}_2$},label=below:$\textbf{\textit{u}}$,right=of t] {$\neg\mathsf{m}$};
\node (x) [label=90:{$\neg\mathsf{e}_1,\mathsf{e}_2$},label=below:$\textbf{\textit{x}}$,right=of u] {$\neg\mathsf{m}$};
\node (w) [label=90:{$\neg\mathsf{e}_1,\mathsf{e}_2$},label=below:$\textbf{\textit{w}}$,right=of x] {$\mathsf{m}$};
\node (z) [label=90:{$e,\neg\mathsf{e}_2$},label=below:$\textbf{\textit{z}}$,right=of w] {$\mathsf{m}$};
\node (s) [label=90:{$\mathsf{e}_1,\mathsf{e}_2$},label=below:$\textbf{\textit{s}}$,right=of z] {$\mathsf{m}$};

\path[<->] (y) edge node[below] {6} (v);

\path[->] (v) edge node[below] {5} (t);

\path[->] (t) edge node[below] {4} (u);

\path[->] (u) edge node[below] {3} (x);

\path[<->] (x) edge node[below] {3} (w);

\path[->] (w) edge node[below] {2} (z);

\path[->] (z) edge node[below] {1} (s);

\end{tikzpicture}
\end{center}

 Suppose now that the officer assigns rating \textbf{D} to the source of $\mathsf{m}$. The source is \textit{not usually reliable}, there is \textit{significant doubt} concerning her honesty even though she has provided valid information in the past. Based on rating \textbf{D}, the officer can mark a \textit{posterior credibility distribution} for message $\mathsf{m}$. Once applied, the scoring rules that correspond to relability rating $\R^\text{D}$ return the following updated model:

\begin{center}
\begin{tikzpicture}[node distance=0.93cm,world/.append style={minimum
size=0.8cm}]
\node (y) [label=90:{$\neg\mathsf{e}_1,\neg\mathsf{e}_2$},label=below:$\textbf{\textit{y}}$,left=of t]  {$\mathsf{m}$};

\node (t) [label=90:{$\mathsf{e}_1,\mathsf{e}_2$},label=below:$\textbf{\textit{t}}$,right=of v] {$\neg\mathsf{m}$};
\node (w) [label=90:{$\neg\mathsf{e}_1,\mathsf{e}_2$},label=below:$\textbf{\textit{w}}$,right=of t] {$\mathsf{m}$};
\node (u) [label=90:{$e,\neg\mathsf{e}_2$},label=below:$\textbf{\textit{u}}$,right=of w] {$\neg\mathsf{m}$};
\node (z) [label=90:{$e,\neg\mathsf{e}_2$},label=below:$\textbf{\textit{z}}$,right=of u] {$\mathsf{m}$};

\node (v) [label=90:{$\neg\mathsf{e}_1,\neg\mathsf{e}_2$},label=below:$\textbf{\textit{v}}$,right=of z] {$\neg\mathsf{m}$};

\node (x) [label=90:{$\neg\mathsf{e}_1,\mathsf{e}_2$},label=below:$\textbf{\textit{x}}$,right=of v] {$\neg\mathsf{m}$};
\node (s) [label=90:{$\mathsf{e}_1,\mathsf{e}_2$},label=below:$\textbf{\textit{s}}$,right=of x] {$\mathsf{m}$};
\node (a) [label=90:{},label=below:,right=of s] {?};

\path[<->] (y) edge node[below] {4} (t);

\path[<->] (t) edge node[below] {4} (w);

\path[->] (w) edge node[below] {3} (u);

\path[<->] (u) edge node[below] {3} (z);

\path[->] (z) edge node[below] {2} (v);

\path[<->] (v) edge node[below] {2} (x);

\path[<->] (x) edge node[below] {2} (s);

\path[->] (s) edge node[below] {1} (a);

\end{tikzpicture}
\end{center}

 For instance, in the initial model, the (prior) credibility rating of message $\mathsf{m}$ was \textbf{2} at state \textit{\textbf{z}}. Now, the \textit{posterior} credibility score of $\mathsf{m}$ is \textbf{3} at state \textit{\textbf{z}}. This suggests that at state \textit{\textbf{z}}, $\mathsf{m}$ is not ``Probably True$"$ (rating \textbf{2}) as it seemed initially but more certainly ``Possibly True$"$ (rating \textbf{3}). Accordingly, the officer should be more careful than she was concerning the truth of message $\mathsf{m}$.   

\medskip
 Table \ref{tab:post} sums up the update operation with rating \textbf{D}. However, two issues can be observed after the computation. Although posterior scores are mutually consistent, rating \textbf{1} is now \textit{undefined}. Besides that, some states, namely $\textit{\textbf{u}}^{\{\mathsf{e}_1,\neg\mathsf{e}_2,\neg\mathsf{m}\}}$ and $\textit{\textbf{z}}^{\{\mathsf{e}_1,\neg\mathsf{e}_2,\mathsf{m}\}}$, are no longer distinguishable. They receive the same credibility score of \textbf{3} that make them indistinguishable. In other words, some credibility information has been lost during the process. But some credibility information has also been gained during the operation. States $\textit{\textbf{x}}^{\{\neg\mathsf{e}_1,\mathsf{e}_2,\neg\mathsf{m}\}}$ and $\textit{\textbf{w}}^{\{\neg\mathsf{e}_1,\mathsf{e}_2,\mathsf{m}\}}$ that were indistinguishable beforehand can now be differentiated. Furthermore, states $\textit{\textbf{v}}^{\{\neg\mathsf{e}_1,\neg\mathsf{e}_2,\neg\mathsf{m}\}}$ and $\textit{\textbf{y}}^{\{\neg\mathsf{e}_1,\neg\mathsf{e}_2,\mathsf{m}\}}$ are no longer evaluative blindspots for the officer since their credibility score is no longer \textbf{6}. 
\medskip

 \begin{table}[!htbp]
  \centering
  
    \begin{tabular}{|c|c|c|}
    \hline
   
    \ \cellcolor{kugray5}\textbf{States} & \cellcolor{kugray5}\textbf{Set of Evidence} & \cellcolor{kugray5}\textbf{Posterior Rating} $\C^{n}\mathsf{m}$ \\
   
    \hline
    
    \textit{\textbf{x}},\textit{\textbf{s}},\textit{\textbf{v}} & $\{\neg\mathsf{e}_1,\mathsf{e}_2\},\{\mathsf{e}_1,\mathsf{e}_2\},\{\neg\mathsf{e}_1,\neg\mathsf{e}_2\}$ & \small{$\neg\C^{2}\mathsf{m}\wedge
   \neg\C^{2}\neg\mathsf{m}$} \ \  \\\hline

  \textit{\textbf{u}},\textit{\textbf{z}} & $\{\mathsf{e}_1,\neg\mathsf{e}_2\}$ & \small{$\neg\C^{3}\mathsf{m}\wedge
   \neg\C^{3}\neg\mathsf{m}$} \ \  \\\hline
    
  \textit{\textbf{t}},\textit{\textbf{w}},\textit{\textbf{y}} & $\{\mathsf{e}_1,\mathsf{e}_2\},\{\neg\mathsf{e}_1,\mathsf{e}_2\},\{\neg\mathsf{e}_1,\neg\mathsf{e}_2\}$ & \small{$\neg\C^{4}\mathsf{m}\wedge
   \neg\C^{4}\neg\mathsf{m}$} \ \

      \\\hline

    \end{tabular}%
\caption{Evidence-Based \textit{Posterior} Ratings.\label{tab:post}}

\end{table}%

We have shown that credibility degrees $\C^{1}$ to $\C^{6}$ and reliability updates $\R^\text{A}$ to $\R^\text{F}$ actually match with credibility ratings from \textbf{1} to \textbf{6} and with reliability ratings from $\textbf{A}$ to $\textbf{F}$. From now on, we use the alphanumeric notation for simplicity's sake, writing simply \textbf{1}, \textbf{2}, \textbf{3}, etc. for  $\C^{1}$, $\C^{2}$, $\C^{3}$, etc., and simply  \textbf{A}, \textbf{B}, \textbf{C} etc. for $\R^\text{A}$, $\R^\text{B}$, $\R^\text{C}$. We propose to consider \textit{all} the posterior scores obtained by applying the reliability updates from \textbf{A} to \textbf{F} to the whole 6$\times$6 landscape of the alphanumeric scale. That being done, we will assume that officers group dimensions into a 3$\times$3 matrix to show that \Logic also adapts well to this context.

\section{Categorizing Intelligence Messages with Scores}
\label{sec:classifying}

\subsection{The Scoring Landscape in the General 6$\times$6 Case}
\label{ssec:scoring}

Our logical procedure agrees with past results on the \textit{credibility versus reliability recommendation}. In accordance with \cite{Baker&al1968}, \cite{Samet1975} and \cite{Phelps&al1980}, \Logic considers credibility as the crucial dimension of evaluation while reliability relies on updating prior credibility degrees. For a given message $\mathsf{m}$ and a set of evidence $\mathsf{E}$ for or against this message, plausibility orderings help determine a \textit{prior credibility distribution} for $\mathsf{m}$. Then, reliability ratings act on this distribution to mark \textit{resultant scores} for $\mathsf{m}$.
\medskip

So far we have not endorsed the assumption that officers group credibility and reliability dimensions into 3$\times$3 categories instead of 6$\times$6 categories. We have left this assumption aside in order to present the most general framework possible applying to the 6$\times$6 dimensions of the alphanumeric scale. Before arguing that our framework is equally well-suited to the 3$\times$3 matrix, Table \ref{tab:landscape} shows the landscape of posterior scores we obtain by applying operations $\textbf{A}$ to $\textbf{F}$ to the prior distribution of credibility ratings \textbf{1} to \textbf{6}, based on all the update rules defined in subsection . 

 \begin{table}[H]
\centering
  \small
{\renewcommand{\arraystretch}{1.4}
\begin{tabular} {|c|c|ccccccccccc|c|} 

 \hline
 
\multicolumn{2}{|c|}{\cellcolor{kugray5}\textbf{Message}} & \multicolumn{12}{c|}{\textit{\textbf{Content Credibility}}}\\

    \cline{3-14}
    \multicolumn{2}{|c|}{\cellcolor{kugray5}\textbf{Scores}} &&\textbf{1}&&\textbf{2}&&\textbf{3}&&\textbf{4}&&\textbf{5}&&\textbf{6}\\
    \hline
    \multirow{6}{*}{\begin{turn}{90}\textit{\textbf{Source Reliability}}\ \ \end{turn}}& \textbf{A}&&1&&1&&1&&1&&2&&1\\
                  & \textbf{B}&&1&&1&&1&&2&&3&&2\\
    
                       & \textbf{C}&&1&&1&&2&&3&&4&&3\\

                      & \textbf{D}&&2&&3&&4&&5&&5&&4\\
    
                     & \textbf{E}&&3&&4&&5&&5&&5&&5\\
    
 \cline{2-14}

       & \textbf{F}&&1&&2&&3&&4&&5&&6\\

    \hline

\end{tabular}}
\caption{General Landscape of Posterior Credibility Scores}
\label{tab:landscape}
\end{table}

Consider for instance a source rated \textbf{D}. Table \ref{tab:landscape} shows that once applied to (prior) credibility rating \textbf{2}, update \textbf{D} returns a posterior credibility score of 3. But once applied to rating \textbf{4}, \textbf{D} returns a posterior credibility rating of 5, etc. Given all the posterior scores of the 6$\times$6 landscape, we may wonder how this landscape adapts to the 3$\times$3 context resulting from the grouping hypothesis \citep{mandel2022meta}. 

\subsection{The Quantitative Ranking of Grouped Ratings}
\label{ssec:meanscores}

We now make the assumption that officers evaluating intelligence messages rely only on a 3$\times$3 scale. Table \ref{tab:landscape-transfo} shows the result of grouping credibility and reliability ratings into three categories, according to observations made by \citep[e.g.][]{Teigen&Brun1995,mandel2022meta}. To obtain Table \ref{tab:landscape-transfo}, we first group the ratings of Table \ref{tab:landscape} based on their directionality. Then, the vertical and horizontal lines corresponding to ratings \textbf{6} and \textbf{F} are centered within the Table, in line with observations of ``neutrality'' for ratings ``Cannot Be Judged'' \cite{mandel2022meta}. Concerning credibility, the uplet (\textbf{1},\textbf{2},\textbf{3}) groups ratings having positive directionality while the uplet (\textbf{4},\textbf{5}) groups ratings having negative directionality. Between those two groups, rating \textbf{6}, whose directionality is neutral (``Cannot Be Judged''), forms an intermediary group. 
Concerning reliability, the uplet (\textbf{A},\textbf{B},\textbf{C}) groups ratings having positive directionality while the uplet (\textbf{D},\textbf{E}) groups ratings having negative directionality. Between those two groups, rating \textbf{F}, whose directionality is neutral (``Cannot Be Judged''), also forms an intermediary group. 

\begin{table}[H]
\centering
  \small
{\renewcommand{\arraystretch}{1.4}
\begin{tabular} {|c|c|ccccccc|c|ccccc|} 

 \hline
 
\multicolumn{2}{|c|}{\cellcolor{kugray5}\textbf{Message}} & \multicolumn{13}{c|}{\textit{\textbf{Content Credibility}}}\\

    \cline{3-15}
    \multicolumn{2}{|c|}{\cellcolor{kugray5}\textbf{Scores}} &&\textbf{1}&&\textbf{2}&&\textbf{3}&&\textbf{6}&&\textbf{4}&&\textbf{5}&\\
    \hline
    \multirow{6}{*}{\begin{turn}{90}\textit{\textbf{Source Reliability}}\ \ \end{turn}}& \textbf{A}&&1&&1&&1&&1&&1&&2&\\
                  & \textbf{B}&&1&&1&&1&&2&&2&&3&\\
    
                       & \textbf{C}&&1&&1&&2&&3&&3&&4&\\

\cline{2-15}
       & \textbf{F}&&1&&2&&3&&6&&4&&5&\\

\cline{2-15}
                      & \textbf{D}&&2&&3&&4&&4&&5&&5&\\
    
                     & \textbf{E}&&3&&4&&5&&5&&5&&5&\\
    
       \cline{2-11}
                   
    \hline

\end{tabular}}
\caption{General Landscape of Posterior Credibility Scores after Grouping}
\label{tab:landscape-transfo}
\end{table}

\medskip
Table \ref{tab:landscape-transfo} is consistent with observations on the groupings of the credibility and reliability dimensions, with ratings \textbf{6} and \textbf{F} standing as intermediary groups. However, as one can see, the numerical value for credibility rating \textbf{6} should be normalized to fit between the values of the group (\textbf{1},\textbf{2},\textbf{3}) and the values of the group (\textbf{4},\textbf{5}). The same operation of normalization should be applied in case of reliability rating \textbf{F} to make the result of this update fit between the result of the grouped updates (\textbf{A},\textbf{B},\textbf{C}) and the result of the grouped updates (\textbf{D},\textbf{E}). 

\medskip
To see more clearly into the link between evaluative scores and descriptive message types, we calculate the means for credibility ratings and for reliability updates, once they are grouped into three categories. Although other calculation rules may be used, calculating average values is the most natural approach when we choose to group scores into uniform categories as it is the case here. From now on, we write ``avg'' the operation which calculates the average, more precisely the mean, for each group of credibility or reliability ratings.  

\medskip
Concerning groups of credibility ratings, we calculate the average of each of the three groupings of credibility ratings. For the positive group (\textbf{1},\textbf{2},\textbf{3}) and the negative group (\textbf{4},\textbf{5}), we obtain avg(\textbf{1},\textbf{2},\textbf{3}) = \textbf{2} and avg(\textbf{4},\textbf{5}) = \textbf{4.5}, respectively. The calculation for avg(\textbf{6}) is slightly different since, according to the grouping hypothesis, credibility rating \textbf{6} is neutral in terms of directionality and stands as an intermediary group between (\textbf{1},\textbf{2},\textbf{3}) and (\textbf{4},\textbf{5}). To comply with this assumption, we define avg(\textbf{6}) as the average between avg(\textbf{1},\textbf{2},\textbf{3}) and avg(\textbf{4},\textbf{5}), that is to say: avg(\textbf{6}) = avg(avg(\textbf{1},\textbf{2},\textbf{3}),avg(\textbf{4},\textbf{5})) = \textbf{3.25}. In doing so, we obtain a normalized score for avg(\textbf{6}) which fits between avg(\textbf{1},\textbf{2},\textbf{3}) and avg(\textbf{4},\textbf{5}).

\medskip
Concerning groups of reliability ratings, Table \ref{tab:updategrouped} provides the average mean rules for each group of reliability updates. In this Table, avg(\textbf{A},\textbf{B},\textbf{C}) is the average reliability update for the positive group (\textbf{A},\textbf{B},\textbf{C}) while avg(\textbf{D},\textbf{E}) is the average reliability update for the negative group (\textbf{D},\textbf{E}). As before, the calculation for avg(\textbf{F}) is slightly different since, according to the grouping hypothesis, \textbf{F} is neutral in terms of directionality and stands as an intermediary group between (\textbf{A},\textbf{B},\textbf{C}) and (\textbf{E},\textbf{F}). To comply with this assumption, we define avg(\textbf{F})($\varphi$) as the average between avg(\textbf{A},\textbf{B},\textbf{C})($\varphi$) and avg(\textbf{E},\textbf{F})($\varphi$). Here avg(\textbf{F}) renormalizes scores contrary to the initial operation \textbf{F} which kept scores as they were initially (see horizontal line for \textbf{F} in Table \ref{tab:landscape-transfo}).

\begin{table}[!htbp]
  \centering
  \begin{tabular}{|c||c|c|}
    \hline
    \rowcolor{kugray5} & & \\
    \rowcolor{kugray5}\textbf{Mean Rating by Group} &  \textbf{If} \textit{\textbf{dg}}(\textit{\textbf{s}}) $\in [1,3.25[\ \cup\  ]3.25,5]$ &  \textbf{If} \textit{\textbf{dg}}(\textit{\textbf{s}}) = 3.25 \\
    \rowcolor{kugray5} &  & \\
    \hline
    & & \\
    avg(\textbf{A},\textbf{B},\textbf{C})($\varphi$)  & \mbox{\normalsize $\textit{\texttt{Cut}}(\textit{\textbf{dg}}(\textit{\textbf{s}}) - 2)$ \textit{if} $\mathds{S}, \textit{\textbf{s}} \models \varphi$} & 2 \textit{if} $\mathds{S}, \textit{\textbf{s}} \models \varphi$ \\ 
    & \mbox{\normalsize $\textit{\texttt{Cut}}(\textit{\textbf{dg}}(\textit{\textbf{s}}) + 2)$ \textit{if} $\mathds{S}, \textit{\textbf{s}} \models \neg\varphi$} & 4 \textit{if} $\mathds{S}, \textit{\textbf{s}} \models \neg\varphi$ \\
    & & \\\hhline{|=|=|=|}
    & \multicolumn{2}{c|}{}  \\
    avg(\textbf{F})($\varphi$) &  \multicolumn{2}{c|}{\normalsize avg(avg(\textbf{A},\textbf{B},\textbf{C})($\varphi$), avg(\textbf{D},\textbf{E})($\varphi$))} \\
    & \multicolumn{2}{c|}{} \\\hhline{|=|=|=|}
    & & \\
    avg(\textbf{D},\textbf{E})($\varphi$)  & \mbox{\normalsize $\textit{\texttt{Cut}}(\textit{\textbf{dg}}(\textit{\textbf{s}}) + 1.5)$ \textit{if} $\mathds{S}, \textit{\textbf{s}} \models \varphi$} & 4.5 \textit{if} $\mathds{S}, \textit{\textbf{s}} \models \varphi$ \\ 
    & \mbox{\normalsize $\textit{\texttt{Cut}}(\textit{\textbf{dg}}(\textit{\textbf{s}}) - 1.5)$ \textit{if} $\mathds{S}, \textit{\textbf{s}} \models \neg\varphi$} & 1.5 \textit{if} $\mathds{S}, \textit{\textbf{s}} \models \neg\varphi$ \\
    &  & \\\hline
  \end{tabular}
  \caption{Means for Grouped Reliability Updates.}
  \label{tab:updategrouped}
\end{table}


\begin{table}[H]
\centering
  \small
{\renewcommand{\arraystretch}{1.4}
\begin{tabular} {|c|c|c|c|c|} 

 \hline
 
\multicolumn{2}{|c|}{\cellcolor{kugray5}\textbf{Average Scores}} & \multicolumn{3}{c|}{\textit{\textbf{Average Content Credibility}}}\\

    \cline{3-5}
    \multicolumn{2}{|c|}{\cellcolor{kugray5}\textbf{by Group}} &\textbf{2}&\textbf{3.25}&\textbf{4.5}\\
    \hline

    \multirow{9}{*}{\begin{turn}{90}\textit{\textbf{Average Source Reliability}} \end{turn}} & & & &\\

&avg(\textbf{A},\textbf{B},\textbf{C})&$1^{\ (i)}$&$2^{\ (v)}$&$2.5^{\ (ii)}$\\
    

                       & &&&\\

\cline{2-5}

  & &&&\\
       & avg(\textbf{F})&$2.25^{\ (vii)}$&$3.25^{\ (ix)}$&$3.75^{\ (viii)}$\\

  & &&&\\

\cline{2-5}

                      & & &&\\
    
      & avg(\textbf{D},\textbf{E}) &$3.5^{\ (iii)}$&$4.5^{\ (vi)}$&$5^{\ (iv)}$\\
                     & & &&\\
    
\cline{2-5}
                   
    \hline

\end{tabular}}
\caption{Posterior Credibility Scores based on Group Means.}
\label{tab:meanscores}
\end{table}


\medskip
In Table \ref{tab:meanscores}, we observe that the average posterior credibility scores are \textit{all distinct} from each others. From a quantitative perspective, mean crediblity scores of groups can be ranked from highest to lowest, as  in the quantitative ranking (\textbf{QT}):

\begin{center}
(\textbf{QT}) \hspace{0.5cm} ($i$) $>_{qt}$ ($v$) $>_{qt}$ ($vii$) $>_{qt}$ ($ii$) $>_{qt}$ ($ix$) $>_{qt}$ ($iii$) $>_{qt}$ ($viii$) $>_{qt}$ ($vi$) $>_{qt}$ ($iv$) 
\end{center}

where ``$>_{qt}$'' means that the numerical average score on the left is \textit{quantitatively} higher that the average score on the right. That being done, we may wonder whether this numerical ranking for average posterior credibility scores actually corresponds to the qualitative ranking of message types established for the taxonomy of intelligence messages proposed in \cite{Icard2023a}. More ambitiously, we may wonder whether the evaluative perspective offered by \Logic converges with the descriptive perspective provided in this taxonomy. We show that this is actually the case: \Logic leads to retrieve the informational types of the taxonomy.

\subsection{Using Evaluative Scores to Identify Descriptive Types of Messages}
\label{ssec:match}

From a theoretical perspective, given an intelligence message $\mathsf{m}$, the \textit{evaluative} credibility score of $\mathsf{m}$ differs from the \textit{descriptive} informational type of $\mathsf{m}$, even though descriptive and evaluative perspectives on information are complementary. As we know, the descriptive type of $\mathsf{m}$ is obtained by crossing the degree of truth of its content with the degree of honesty of its source. By contrast, the evaluative score of $\mathsf{m}$ is obtained by marking a posterior credibility score for $\mathsf{m}$ based on its prior credibility score updated by its source's reliability.
\medskip

As one can see, the quantitative ranking (\textbf{QT}) is perfectly congruent with the qualitative ranking (\textbf{QL}) inferred from the 3$\times$3 taxonomy of intelligence messages (see subsection \ref{ssec:informalicard}). This correspondence matches with intuitions: the more beneficial the message is based on its informational type, the better its resultant score. For intance, $\textbf{\textit{\texttt{t}}}_{\textbf{\texttt{\textit{1}}}}$ = \textit{information} and $\textbf{\textit{\texttt{t}}}_{\textbf{\texttt{\textit{2}}}}$ = \textit{error-avoidance} are the best informational types in (\textbf{QL}) and, accordingly, the average score is 1 for (\textit{i}) in (\textbf{QT}) while the average score is 2 for (\textit{ii}) in (\textbf{QT}). By contrast, the more harmful or detrimental the message is with respect to its type, the worse its resultant score. Now, $\textbf{\textit{\texttt{t}}}_{\textbf{\texttt{\textit{4}}}}$ = \textit{objective lie} and $\textbf{\textit{\texttt{t}}}_{\textbf{\texttt{\textit{6}}}}$ = \textit{half-truth} are the worst informational types in (\textbf{QL}) and, accordingly, the average score is 5 for (\textit{iv}) in (\textbf{QT}), while the average score is 4.5 for (\textit{vi}) in (\textbf{QT}). This qualitative-quantitative correspondence is also valid for all the other types of the taxonomy.

   \begin{table}[H]
\centering
  \small
{\renewcommand{\arraystretch}{1.4}
\begin{tabular} {|c|c|ccccccc|c|ccccc|} 

 \hline
 
\multicolumn{2}{|c|}{\cellcolor{kugray5}\textbf{Message}} & \multicolumn{13}{c|}{\textit{\textbf{Content Credibility}}}\\

    \cline{3-15}
    \multicolumn{2}{|c|}{\cellcolor{kugray5}\textbf{Scores}} &&\textbf{1}&|&\textbf{2}&|&\textbf{3}&&\textbf{6}&&\textbf{4}&|&\textbf{5}&\\
    \hline
    \multirow{12}{*}{\begin{turn}{90}\textit{\textbf{Source Reliability}} \end{turn}}& \textbf{A}&&\textbf{A1}&|&\textbf{A2}&|&\textbf{A3}&&\textbf{A6}&&\textbf{A4}&|&\textbf{A5}&\\
    \cline{2-2}
                  & \textbf{B}&&\textbf{B1}&|&\textbf{B2}&|&\textbf{B3}&&\textbf{B6}&&\textbf{B4}&|&\textbf{B5}&\\
    
       \cline{2-2}
                       & \textbf{C}&&\textbf{C1}&|&\textbf{C2}&|&\textbf{C3}&&\textbf{C6}&&\textbf{C4}&|&\textbf{C5}&\\

    & & \multicolumn{7}{c|}{$\textbf{\textit{\texttt{t}}}_{\textbf{\texttt{\textit{1}}}}\equiv$ (\textit{i})} &\textit{$\textbf{\textit{\texttt{t}}}_{\textbf{\texttt{\textit{5}}}}\equiv (\textit{v})$}&\multicolumn{5}{c|}{$\textbf{\textit{\texttt{t}}}_{\textbf{\texttt{\textit{2}}}}\equiv (\textit{ii})$}\\

& & \multicolumn{7}{c|}{\textit{information}} &\textit{error-avoidance}&\multicolumn{5}{c|}{\textit{misinformation}}\\

\cline{2-15}
       & \textbf{F}&&\textbf{F1}&|&\textbf{F2}&|&\textbf{F3}&&\textbf{F6}&&\textbf{F4}&|&\textbf{F5}&\\

       & & \multicolumn{7}{c|}{$\textbf{\textit{\texttt{t}}}_{\textbf{\texttt{\textit{7}}}} \equiv (\textit{vii})$} &$\textbf{\textit{\texttt{t}}}_{\textbf{\texttt{\textit{9}}}} \equiv (\textit{ix})$&\multicolumn{5}{c|}{$\textbf{\textit{\texttt{t}}}_{\textbf{\texttt{\textit{8}}}} \equiv (\textit{viii})$}\\

& & \multicolumn{7}{c|}{\textit{omission}} &\textit{mixed}&\multicolumn{5}{c|}{\textit{dissimulation}}\\

\cline{2-15}
                      & \textbf{D}&&\textbf{D1}&|&\textbf{D2}&|&\textbf{D3}&&\textbf{D6}&&\textbf{D4}&|&\textbf{D5}&\\
    
       \cline{2-2}
                     & \textbf{E}&&\textbf{E1}&|&\textbf{E2}&|&\textbf{E3}&&\textbf{E6}&&\textbf{E4}&|&\textbf{E5}&\\

       & & \multicolumn{7}{c|}{$\textbf{\textit{\texttt{t}}}_{\textbf{\texttt{\textit{3}}}} \equiv (\textit{iii})$} &$\textbf{\textit{\texttt{t}}}_{\textbf{\texttt{\textit{6}}}} \equiv (\textit{vi})$&\multicolumn{5}{c|}{$\textbf{\textit{\texttt{t}}}_{\textbf{\texttt{\textit{4}}}} \equiv (\textit{iv})$}\\

  & & \multicolumn{7}{c|}{\textit{subjective lie}} &\textit{half-truth}&\multicolumn{5}{c|}{\textit{objective lie}}\\                   
       \cline{2-11}
                   
    \hline

\end{tabular}}
\caption{Connecting Descriptive and Evaluative Perspectives on the Grouped 6$\times$6 Matrix.}
\label{tab:superposition}
\end{table}

In Table \ref{tab:superposition}, we summarize the correspondence between the descriptive taxonomy of intelligence proposed by \cite{Icard2023a} (Table \ref{tab:icardtaxo}) and the mean credibility scores obtained with \Logic (Table \ref{tab:meanscores}). An equivalence symbol, written $\equiv$, shows the connection between those two proposals which consist in giving descriptive versus evaluative insights on intelligence messages (for instance, $\textbf{\textit{\texttt{t}}}_{\textbf{\texttt{\textit{6}}}} \equiv (\textit{vi})$ indicates the correspondence between descriptive type $\textbf{\textit{\texttt{t}}}_{\textbf{\texttt{\textit{6}}}}$ and the evaluative mean score (\textit{vi})). 
In each case, we also provide the linguistic labels associated to the messages in the descriptive taxonomy (for instance, \textit{half-truth} for $\textbf{\textit{\texttt{t}}}_{\textbf{\texttt{\textit{6}}}} \equiv (\textit{vi})$).

\section{Conclusion}
\label{sec:conclusion}







Experimental studies on information evaluation resulted in several observations. First of all, officers do not have a clear understanding of the 6$\times$6 Admiralty scale. They tend to use only a 3$\times$3 subpart of this scale by grouping the credibility and reliability dimensions in 3 categories instead of 6. Second, officers are unable to respect the recommendation to distinguish facts from interpretations with the current 6$\times$6 scale, which does not allow them to give a descriptive insight on intelligence messages. Third, officers cannot respect the recommendation to distinguish credibility from reliability because they actually perceive those dimensions as strongly correlated. More specifically, they consider credibility as the most important dimension during information evaluation while reliability only plays an adjustement role.

\medskip
In this article, we adaptated extant work in  dynamic belief revision to make credibility the main dimension of evaluation \citep[e.g.][]{Aucher2004,vanDitmarsch2005,VanDitmarsch&Labuschagne2007}. In the framework \Logic, we rely on numerical plausibility models to define credibility operators from \textbf{1} to \textbf{6}. Then, we use dynamic reliability operators from \textbf{A} to \textbf{F} to update those plausibility models depending on the reliability of the source. To keep things general, we only assumed that credibility was the main dimension of evaluation and that reliability played an adjustment role. We did not make any other assumption, for instance that officers actually use only 3$\times$3 dimensions of the alphanumeric scale. However, we have shown that \Logic can also be adapted to three dimensions for credibility and reliability instead of six. In this case, the 3$\times$3 version of \Logic can be used to retrieve, and ascribe, the taxonomy of intelligence messages proposed by \cite{Icard2023a}, the purpose of which being to help officers better separate facts from interpretations in intelligence.    

\medskip
Our proposal aims to bridge the gap between qualitative and quantitative approaches to information evaluation. We expect our approach to be of practical use within the intelligence community by corresponding to the way officers actually deal with the credibility and reliability dimensions. As a formal procedure, \Logic could be implemented computationnaly to help classify messages according to more explicit and significant categories (viz. \textit{information}, \textit{misinformation}, \textit{lies}, etc.). But the logic is general enough to extend outside intelligence. It can apply to  informational contexts also strongly concerned by the evaluation of sources and the identification of \textit{biased} or \textit{fake} information (as in news press for instance).

\medskip 
We consider two complementary directions for future work. As a first step, we aim to see more clearly into the logical interactions between pieces of evidence and contents of intelligence messages. In the current \Logic, contents and evidence are atomic formulas which receive the same semantic interpretation. This interpretation should be refined to express the semantic difference between evidence and contents and to weigh the support, or lack of support, evidence provides to contents during information evaluation. As a second step, we aim to see whether this more expressive version of \Logic applies on a wider scope, to other information processing tasks. 

\medskip
Information evaluation is just one aspect of information processing, the other aspect being intelligence analysis through the marshaling of hypotheses. Intelligence analysts evaluate the plausibility of hypotheses in a broader context, based on evidence obtained by other gathering techniques, from other sensor types. Since analysts usually rank hypotheses by considering evidence consistent or inconsistent with their plausibility, the way they proceed is similar to the way intelligence officers proceed, at least to some extent. For this reason, we aim to determine whether the logic we have defined for information evaluation extend to intelligence analysis with some technical adjustments.

\section*{Acknowledgments}
Funding support for this work was provided by project DIEKB (DGA01D19018444), the ANR program HYBRINFOX (ANR-21-ASIA-0003) and the ANR program Front\-Cog (ANR-17-EURE-0017). 

\newpage
\bibliographystyle{elsarticle-num-names} 
\bibliography{lib}

\end{document}